\documentclass{aa}
\usepackage{graphicx}
\usepackage[varg]{txfonts}
\usepackage{physics}
\usepackage{dsfont}
\usepackage{hyperref}

\newcommand{\sref}[1]{Sec.~\ref{#1}}
\newcommand{\tab}[1]{Table~\ref{#1}}
\newcommand{\fig}[1]{Fig.~\ref{#1}}
\newcommand{\figl}[1]{Figure~\ref{#1}}
\newcommand{\equ}[1]{Eq.~(\ref{#1})}
\newcommand{\equs}[2]{Eqs.~(\ref{#1})~-~(\ref{#2})}

\newcommand{\pder}[2][]{\frac{\partial#1}{\partial#2}}

\usepackage{color}

\begin{document}

\titlerunning{The importance of stellar magnetic torques for the inner disk regions.}
\authorrunning{D. Steiner et al.}
\title{Time-dependent, long-term hydrodynamic simulations of the inner protoplanetary disk I: The importance of stellar magnetic torques}
\author{
 D.~Steiner\inst{1},
 L.~Gehrig\inst{1},
 B.~Ratschiner\inst{1},
 F.~Ragossnig\inst{1},
 E.~I.~Vorobyov\inst{1,2},
 M.~Güdel\inst{1},
 E.~A.~Dorfi \dag \inst{1}
}
\institute{
 Institute for Astronomy (IfA), University of Vienna,
 Türkenschanzstrasse 17, A-1180 Vienna
 \and
 Institute of Astronomy, Russian Academy of Sciences,
 48 Pyatnitskaya St., Moscow, 119017, Russia
}
\date{Received ... / Accepted 06-21-2021}

\abstract
{}
{We conduct simulations of the inner regions of protoplanetary disks (PPDs) to investigate the effects of protostellar magnetic fields on their long-term evolution. We use an inner boundary model that incorporates the influence of a stellar magnetic field. The position of the inner disk is dependent on the mass accretion rate as well as the magnetic field strength. We use this model to study the response of a magnetically truncated inner disk to an episodic accretion event. Additionally, we vary the protostellar magnetic field strength and investigate the consequences of the magnetic field on the long-term behavior of PPDs.}
{
We use the fully implicit 1+1D TAPIR code which solves the axisymmetric hydrodynamic equations self-consistently. Our model allows us to investigate disk dynamics close to the star and to conduct long-term evolution simulations  simultaneously. We assume a hydrostatic vertical configuration described via an energy equation which accounts for the radiative transport in the vertical direction in the optically thick limit and the equation of state. Moreover, our model includes the radial radiation transport in the stationary diffusion limit and takes protostellar irradiation into account.
}
{We include stellar magnetic torques, the influence of a pressure gradient, and a variable inner disk radius in the TAPIR code to describe the innermost disk region in a more self-consistent manner. We can show that this approach alters the disk dynamics considerably compared to a simplified diffusive evolution equation, especially during outbursts. During a single outburst, the angular velocity deviates significantly from the Keplerian velocity because of the influence of stellar magnetic torques. The disk pressure gradient switches sign several times and the inner disk radius is pushed towards the star, approaching $< 1.2$ $R_\mathrm{\star}$. Additionally, by varying the stellar magnetic field strength, we can demonstrate several previously unseen effects. The number, duration, and the accreted disk mass of an outburst as well as the disk mass at the end of the disk phase (after several million years) depend on the stellar field strength. Furthermore, we can define a range of stellar magnetic field strengths, in which outbursts are completely suppressed. The robustness of this result is confirmed by varying different disk parameters.}
{The influences of a prescribed stellar magnetic field, local pressure gradients, and a variable inner disk radius result in a more consistent description of the gas dynamics in the innermost regions of PPDs. Combining magnetic torques acting on the innermost disk regions with the long-term evolution of PPDs yields previously unseen results, whereby the whole disk structure is affected over its entire lifetime. Additionally, we want to emphasize that a combination of our 1+1D model with more sophisticated multi-dimensional codes could improve the understanding of PPDs even further.}


\maketitle

\section{Introduction}
\label{sec:intro}

During the collapse of a molecular cloud, a protostar and its surrounding protostellar disk are formed. Observations of such star--disk systems reveal that the luminosity of the central object varies with time \citep[e.g.][]{herbig89, contreras17}. Some of those young variable stars can be classified as FU Orionis (FU Ori) objects, which show a rise in luminosity over several orders of magnitudes for a short period of time \citep[e.g.,][]{hartmann96, audard2014}. Spectroscopic observations by \citet{zhu07} and \citet{eisner11} indicate that such high-luminosity phases (bursts) are associated with an increased accretion rate of disk material onto the protostar. The frequency of FU Ori-type eruptions in the solar neighborhood suggests that a protostellar accretion burst is not a single but a recurring phenomenon \citep{hartmann96}. Moreover, numerical simulations indicate that an FU Ori-like star--disk system undergoes at least 10 to 20 such eruptive events \citep{hartmann96}. 

Thermal instability (TI) of the inner disk provides a possible explanation for the aforementioned eruptive episodic accretion events. Ionization of most of the hydrogen in the very inner parts of a disk ($10 \, R_{\odot} \lesssim r \lesssim 0.1 \rm AU$) leads to enhanced viscosity caused by runaway heating of the optically thick inner disk regions and therefore to an increased accretion rate \citep[e.g.,][]{bell94}. Although their model is able to predict bursts, the explanation of the burst duration is unsatisfying because of the narrow radial region in which the disk can become sufficiently ionized and hence may become thermally unstable \citep[][]{zhu09}. To expand the potentially thermally unstable regions of a disk, a model with vertical layers of different viscosity has been proposed \citep[e.g.,][]{gammie96}. A warm, irradiated surface layer is sufficiently ionized for the magneto-rotational instability (MRI) to efficiently operate, but also acts as a shield for the deeper disk layer close to the midplane. Hence, MRI cannot develop  in the deep layer and the subsequent reduced viscosity leads to a pile-up of material from the outer disk (dead zone). Eventually, the gas temperature in the dead zone rises above a certain threshold, caused by enhanced viscous heating of the accumulated mass. The deep disk layer then becomes MRI unstable, which may be sufficient to trigger the TI in the dead zone, effectively broadening the radial region where the disk can become thermally unstable. Recent work \citep[e.g.,][]{armitage01, zhu09, zhu10} assumed gravitational instabilities (GIs) to act as an effective mechanism to enhance the mass transport rate from the outer disk towards the inner regions. However, observations of low-luminosity accretion events ---identified as FU Ori events \citep[see e.g.,][]{kospal16, hillenbrand18}--- show that low-mass protostellar star--disk systems with masses lower than  $0.01 M_{\odot}$ can also undergo episodic accretion. Those systems are, contrary to  FU Ori for example, not in the embedded (Class I) but in the late Class II phase \citep[cf.][]{kospal16} and additionally have disks with masses $M_{\rm disk} \approx 0.01\, M_{\odot}$, which are too low for GI to contribute significantly (even in the outer disk) to the averaged mass transport rate.

As TI is likely to develop in the very inner disk parts, especially for low-mass disks, the position of the inner disk rim is a crucial parameter. Magneto-hydrodynamic (MHD) simulations in 3D show that protoplanetary disks are eventually disrupted by strong protostellar magnetic fields \citep[e.g.,][]{Romanova04, bouvier07, romanova14, Romanova15}. The accretion flow is then funneled along magnetic field lines onto the star. This is especially important in the case of episodic accretion, as then the inner disk rim is moving towards the star \citep[e.g.,][]{hartmann16, zhu19} and consequently the inner disk dynamics is altered. To be able to consider the influence of stellar magnetic torques on the disk evolution and simultaneously calculate the radial position of the inner disk radius  self-consistently, it is necessary to solve the full set of hydrodynamic equations.

The work of  \citet{bell94}, \citet{armitage01} and \citet{zhu07, zhu08, zhu09} for example is able to explain FU Ori-like outbursts by either purely thermally unstable inner disks or by including a layered viscosity model. Due to time-step limitations of explicit methods, especially in disk regions close to the star ($r \leq 0.5$ AU) (Appendix~\ref{sec:imp_timestep}), neither the influence of protostellar magnetic fields nor the effects of local steep pressure gradients could be reproduced. However, the findings of~\citet{bouvier07}~and~\citet{romanova14}, for example, indicate that the very inner regions of disks are crucial for understanding episodic accretion. Recent hydrodynamic models of MRI and thermal instability bursts \citep[][]{bae2013, kadam20, vorobyov20}) have strong limitations in modeling the innermost disk regions, therefore necessitating the development of numerical codes that can realistically treat the star--disk interface in global disk simulations. Combining the requirements of (i) self-consistent treatment of both the gas dynamics in the inner disk including the effects of a stellar magnetic field and the position of the inner disk radius, (ii) the inclusion of an energy equation to investigate the effects of TI-induced episodic accretion, and (iii) maintaining the ability to resolve the very inner disk parts of a few stellar radii and carry out long-term calculations up to a few million years, we choose to use an implicit 1+1D RHD code \citep[TAPIR, see][]{stoekl14a, ragossnig20} that enables us to meet all the requirements without being time-step-limited by the Courant-Friedrichs-Levy (CFL) condition (Appendix~\ref{sec:imp_timestep}). Until now, no comparable numerical code has combined the above requirements. We also compare our simulations to similar work \citep[e.g.,][]{zhu10a, zhu10b, zhu19} and point out the differences due to a time-dependent and self-consistent treatment of the component $u_{\varphi}(r, t)$.
 
In \sref{sec:physical_setup} we briefly describe the physical equations and the layered viscosity model. Details of our method and the implementation of the boundary conditions used are explained in \sref{sec:method_description}.
The position of the inner disk edge as well as the importance of a consistent treatment of the inner disk region is outlined in \sref{sec:inner_boundary}. Our results are presented and discussed in \sref{sec:results} and are then summarized in \sref{sec:conclusion}.

\section{Physical setup}
\label{sec:physical_setup}

Protoplanetary disks and their evolution in time can be described by adopting the equations of hydrodynamics using the turbulent viscosity prescription introduced by for example~\citet{Shakura1973} or~\citet{balbus91} and an energy equation. This set of equations is briefly presented in Section~\ref{sec:theoretical_framework}. Additionally, the stellar magnetic field (Sec.~\ref{sec:stellar_magnetic_field}), a description of the energy equation (Sec.~\ref{sec:energy_equation}), the adapted viscosity model (Sec.~\ref{sec:visco_model}), and the gas and dust opacities (Sec.~\ref{sec:opacities}) are presented.

\subsection{Theoretical framework}
\label{sec:theoretical_framework}

The following equations describe the time evolution of a viscous accretion disk in the thin-disk limit (pressure scale-height $H_p \ll$ radius $r$) around a protostar. The energy equation contains radial and vertical radiation transport in an approximated way (see \sref{sec:energy_equation}) and incorporates irradiation from the central object. Because of the thin-disk approximation, the equations can be vertically integrated and read as,
\begin{alignat}{2}
    & \pder{t} \, \Sigma &&+ \nabla \cdot ( \Sigma \, \vec u ) = 0\;, \label{eq:cont} \\
    & \pder{t} (\Sigma \, \vec u) &&+ \nabla \cdot (\Sigma \, \vec u : \vec u) - \frac{B_\mathrm{z} \vec B}{2 \pi}\nonumber \\
        & &&+ \nabla P_\mathrm{gas} + \nabla \cdot Q + \Sigma \, \nabla \psi + H_\mathrm{p} \, \nabla \left( \frac{B_\mathrm{z}^2}{4 \pi} \right) = 0 \;, \label{eq:mot} \\
    &\pder{t} (\Sigma \, e) &&+ \nabla \cdot (\Sigma \, \vec u \, e ) + P_\mathrm{gas} \, \nabla \cdot  \vec u \nonumber \\
    & &&+ Q : \nabla \vec u - 4 \pi \, \Sigma \, \kappa\left(J - S \right) + \dot E_\mathrm{rad} = 0 \;, \label{eq:ene}
\end{alignat}

where $\Sigma$, $P_\mathrm{gas}$, and $\vec u$ denote the gas column density, the vertically integrated gas pressure, and the gas velocity vector, respectively. $Q$ represents the viscous pressure tensor and $\psi$ marks the gravitational potential of the system (star and disk). In most of our calculations, we focus on late class II systems, in which the gravitational force from the protostar dominates. The protostellar magnetic field is presented by $\vec B = (B_\mathrm{r}, B_\mathrm{\varphi}, B_\mathrm{z})^T$ and is modeled via a prescribed field in polodial and torodial directions (see \sref{sec:stellar_magnetic_field}). The term $B_\mathrm{z} \, \vec B / 2 \, \pi$ corresponds to magnetic stress exerted on the disk gas, whereas $\nabla(B_\mathrm{z}^2 / 4 \pi )$ describes the magnetic pressure force density. In \equ{eq:ene}, $e$ denotes the specific internal energy density, $J$ stands for the zeroth moment of the radiation field and $S$ represents the source function. The $(J - S)$ in \equ{eq:ene} is treated approximately as described in \sref{sec:energy_equation}. To describe the internal structure of the ideal gas in radial and vertical directions, we utilize the ideal equation of state (EOS),
\begin{align}
    P_\mathrm{gas} = ( \gamma - 1 ) \, e \;,
\end{align}
where $\gamma $ denotes the adiabatic coefficient.

\subsection{Stellar magnetic field}
\label{sec:stellar_magnetic_field}

\citet{livio92} used an analytic approach to describe a stellar magnetic field which threads an accretion disk, 
\begin{align}
    B_\mathrm{z}(r) &= B_\star \left(\frac{R_\star}{r}\right)^3 \;, \label{eq:stellar_Bz} \\
    B_\mathrm{\varphi}^I(r) &\simeq B_\mathrm{z}(r) \left( 1- \frac{\Omega(r)}{\Omega_\star} \right) \;, \label{eq:stellar_Bphi_livio}
\end{align}
where $B_\mathrm{z}$ is the vertical magnetic field component and is assumed to have a positive sign. The stellar magnetic field is modeled in \equs{eq:stellar_Bz}{eq:stellar_Bphi_livio} as a dipole field and a toroidal component $B_\mathrm{\varphi}^I$ due to the torque exerted on the threaded protoplanetary disk by the stellar magnetic field, respectively. However, \equ{eq:stellar_Bphi_livio} has the problem that $|B_\mathrm{\varphi}| \gg |B_\mathrm{z}|$ very close to the star. Therefore, various authors \citep[e.g.,][]{rappaport04, kluzniak07} use a slightly modified version of \equ{eq:stellar_Bphi_livio}, which reads,
\begin{alignat}{2}
    & \alpha_\mathrm{cor} &&=
    \begin{cases}
      +1 & \text{if $r < r_\mathrm{cor}$} \\
      -1 & \text{if $r > r_\mathrm{cor}$} \\
    \end{cases} \;, \\
    & B_\mathrm{\varphi}^{II}(r) &&\simeq - \alpha_\mathrm{cor} \, B_\mathrm{z}(r) \left[ 1- \left(\frac{\Omega_\star}{\Omega(r)}\right)^{\alpha_\mathrm{cor}} \right] \;. \label{eq:stellar_Bphi}
\end{alignat}
The superscripts $I$ and $II$ are applied to differentiate between the two models for $B_\mathrm{\varphi}$, whereas $\alpha_\mathrm{cor}$ takes account of changing the field prescription to inside and outside the corotation radius. The model described by $B_\mathrm{\varphi}^{II}$ confers the advantage that $|B_\mathrm{\varphi}| \approx |B_\mathrm{z}|$ for radii not close to $r_\mathrm{cor}$. This is also physically motivated, as larger $|B_\mathrm{\varphi}|$ are not possible because a too tightly wound up toroidal field would lead to an opening up of the field configuration because of the increased magnetic energy \citep[see e.g.,][]{rappaport04}. Typical values for the protostellar magnetic field are of the order of several kG \citep[see e.g.,][]{bouvier07, Kurosawa08}.

\subsection{Energy equation}
\label{sec:energy_equation}

The energy equation \equ{eq:ene} is described by \citet{ragossnig20} and accounts for viscous heating, radiative cooling, and radial diffusion of inner energy. Furthermore, a vertical radiation transport is included to also account for stellar irradiation on to the surface of the disk. In cylindrical coordinates, \equ{eq:ene} reads,

\begin{align}
    &Q = \frac{\mu}{2} \left[ \nabla \vec{u} + (\nabla \vec{u})^T - \frac{2}{3}(\nabla \cdot \vec{u})~\mathds{1} \right] \;, \label{eq:shearstresstensor} \\
    &\epsilon_\mathrm{Q} \equiv 2 \, Q : {\vec \nabla} {\vec u} \; ,
\end{align}
\begin{alignat}{2}
    &\pder{t} \left( \Sigma \, e \right)
        &&+ \frac{1}{r}\pder{r} \left( r \, u_\mathrm{r} \, \Sigma \, e \right) 
        + P \frac{1}{r}\pder{r} \left( r \, u_\mathrm{r} \, \right) 
        + \epsilon_Q \nonumber  + \dot E_\mathrm{rad} \\ 
        & &&- 4\pi \, \frac{\sigma}{3 \, r \, \pi}\pder{r}\left( \frac{r}{\kappa_\mathrm{R} \, \rho} \pder[T_0^4]{r} \right) = 0 \;,  \label{eq:ene_cyl}
\end{alignat}
where the second and third terms in \equ{eq:ene_cyl} represent advection due to accretion and work done by pressure, respectively. Here, $\epsilon_Q$ denotes the viscous energy dissipation \citep[e.g.,][]{Tscharnuter1979}. The last term in \equ{eq:ene_cyl} describes radial radiative transport as a diffusion approximation in the Eddington limit \citep[e.g.,][]{ragossnig20}, whereas $\dot E_\mathrm{rad}$ depicts the net radiation heating or cooling rate per unit surface area of the protoplanetary disk. This latter can be obtained by solving the vertical, stationary radiation transfer equation in the optically thick limit ($\tau \rightarrow \infty$) assuming local thermal equilibrium and utilizing an Eddington factor $f_\mathrm{edd} = 1/3$,
\begin{align}
   \dot E_\mathrm{rad} &= \frac{8 \sigma}{3 \tau} \left( T_\mathrm{0}^4 - T_\mathrm{surf}^4 \right) \label{eq:rad_transfer} \:.
\end{align}
$\dot E_\mathrm{rad}$ is a balance of protostellar irradiation $\dot E_\mathrm{irr}$, radiative cooling $\dot E_\mathrm{cool}$, and irradiation from the ambient medium $\dot E_\mathrm{amb}$,
\begin{alignat}{2}
    &\dot E_\mathrm{rad} &&+ \dot E_\mathrm{irr} + \dot E_\mathrm{amb} - \dot E_\mathrm{cool} = 0 \:, \\
    &\dot E_\mathrm{irr}  &&= \frac{L_\star}{r^2 \, \pi} \, f_\mathrm{irr} \, \max \left[ \Delta{\left( \frac{H_p - H_\star}{r} \right)},0 \right] \:, \\
    &\dot E_\mathrm{amb}  &&= 2 \, \sigma T_\mathrm{amb}^4 \:, \\
    &\dot E_\mathrm{cool} &&= 2 \, \sigma T_\mathrm{surf}^4 \:,
\end{alignat}
where $L_\star$ and $H_\star$ denote stellar luminosity and an effective stellar radius, respectively. Furthermore, black-body irradiation is assumed for the cooling term and the ambient radiation. The final form reads \citep[][]{ragossnig20},
\begin{alignat}{2}
        &\dot E_\mathrm{rad} &&=  \sigma \frac{1}{1+\tau'} \left( T_0^4 - T_\mathrm{amb}^4 \right)
        -\frac{L_\star}{r^2 \, \pi} \, f_\mathrm{irr} \, \max \left[ \Delta{\left( \frac{H_p - H_\star}{r} \right)}, \,0 \right] \;, \\
        &\tau' &&\equiv \frac{3}{4} \tau \:, 
        \label{eq:E_rad_final}
\end{alignat}
where $f_\mathrm{irr}$ determines how much of the irradiation can be processed by the gas and which fraction gets reflected.

\subsection{Viscosity model}
\label{sec:visco_model}

The kinematic viscosity $\nu$ is defined as \citep{Shakura1973}
\begin{equation}
    \nu = \alpha \, c_\mathrm{S} \,  H_\mathrm{P} \;, \label{eq:shakura}
\end{equation}
where $\alpha$ is the viscous parameter, $c_S$ the isothermal sound speed, and $H_\mathrm{p}$ the pressure scale height. 
We apply the thin-disk approximation \citep[][]{armitage10} which allows a vertical integration of the physical disk quantities via the EOS, 
\begin{equation}
    P_\mathrm{gas, 0} = \rho_\mathrm{0} \, c_\mathrm{S}^2 \;,
    \label{eq:eos_midplane}
\end{equation}
where  $P_\mathrm{gas, 0}$ is the midplane gas pressure. The vertically integrated dynamical viscosity $\mu$ is then written as 
\begin{equation}
    \mu = \alpha \, c_{\mathrm{S}} \, H_{\mathrm P} \, \Sigma  \;. \label{eq:mu}
\end{equation}
For the viscosity parameter $\alpha$ we adopt the layered-viscosity model \citep{gammie96}, which is a sum of contributions of vertically stratified disk layers, 
\begin{equation}
    \alpha = \alpha_\mathrm{base} + \alpha_\mathrm{surf} + \alpha_\mathrm{deep} + \alpha_\mathrm{grav} \,, \label{eq:alpha}
\end{equation}
where $\alpha_\mathrm{base}$ is a base value for the viscosity (e.g., within a dead-zone) and $\alpha_\mathrm{surf}$ is the MRI viscosity in the permanently active surface layer, which is ionized by cosmic rays and stellar irradiation  for example. $\alpha_\mathrm{deep}$ accounts for the viscosity in parts of the disk, where the temperature exceeds the threshold for thermal ionization \citep[e.g.,][]{zhu09}. The parameter $\alpha_\mathrm{grav}$ denotes GIs acting as effective viscosity.

The separation of the surface and deep layer is controlled by the local surface density. If the surface density at an arbitrary radius exceeds a value $\Sigma_\mathrm{0}$, the disk develops a deep disk layer at that radius. To handle this, we introduce a density switch that is $s_\Sigma = 1$ if $\Sigma \geq \Sigma_\mathrm{0}$ and $s_\Sigma = 0$ if $\Sigma \leq \Sigma_\mathrm{0}$. Parts of the deep layer then become MRI active if the local gas temperature $T_0$ is higher than a constant temperature threshold $T_\mathrm{active}$ for thermal ionization. Hence, the viscosity parameter for the surface layer reads
\begin{equation}
    \alpha_\mathrm{surf}(r) = \alpha_\mathrm{MRI} \left[ s_\Sigma \, \frac{\Sigma_\mathrm{0}}{\Sigma(r)}+(1-s_\Sigma) \right] \;,
\end{equation}
where $\alpha_\mathrm{MRI}$ \citep[e.g.,][]{zhu10, hartmann18} is a constant value for MRI-active regions. We adopt  $\Sigma_\mathrm{0} = 100 \, \mathrm{g \, cm^{-2}}$ for all calculations in this work. Additionally, the viscosity parameter for the deep layer is temperature dependent, that is,
\begin{equation}
    \alpha_\mathrm{deep}(r) = \alpha_\mathrm{MRI} \, s_\Sigma \left(1 - \frac{\Sigma_\mathrm{0}}{\Sigma(r)} \right) s\left( T_0 \right) \;,
\end{equation}
where $s\left( T_0 \right)$ is a temperature switch, 
\begin{equation}
    s\left( T_0 \right) = \frac{1}{2} \left[ 1 + \tanh \left( \frac{T_0 - T_\mathrm {active}}{T_\mathrm{width}} \right) \right] \;,
    \label{eq:smooth}
\end{equation}
which controls whether the deep layer is MRI-active or not \citep[see e.g.,][]{flock16}. For faster convergence of the Newton-Raphson iteration method (see \sref{sec:method_description}), \equ{eq:smooth} is utilized to establish a smooth transition between active and inactive parts within the deep layer, where $T_\mathrm{width} = 10 - 50~\rm K$ is a smoothing width for the temperature. This smooth transition creates a temperature zone below $T_\mathrm{active}$ in which viscosity is already increased and the disk becomes MRI unstable. This MRI active zone starts at $\approx T_\mathrm{active} - 2 \cdot T_\mathrm{width}$.

We define the viscosity parameter for gravitationally unstable regions as
\begin{equation}
    \alpha_\mathrm{grav} = \alpha_\mathrm{GI} \, s_\mathrm{G} \left( \frac{Q^2_\mathrm{T, crit.}}{Q^2_\mathrm{T}}-1 \right) \,,
\end{equation}
where $\alpha_\mathrm{GI} = 0.01$ is a constant value for gravitationally unstable regions within the disk, and $Q_\mathrm{T}$ is the Toomre parameter \citep{toomre64}.
The critical value for the Toomre parameter at which GIs occur is set to $Q_\mathrm{T, crit} = 1.0$. The quantity $s_\mathrm{G}$ is a switch for the GI and is $s_\mathrm{G} = 1$ if the disk is unstable ($Q_\mathrm{T} < Q_\mathrm{T, crit.}$) and $s_\mathrm{G} = 0$ otherwise. In this paper we have chosen disks with masses low enough ($M_{disk} < 0.1 M_\star$) to not exceed the Toomre criterion, thus avoiding GIs within the disk. 
%
\subsection{Gas and dust opacities}
\label{sec:opacities}
For the gas opacity description $\kappa_\mathrm{R, gas}$ we use the Rosseland-mean opacity tables created by \citet{ferguson05}, who compiled low-temperature opacity tables by employing opacity sampling methods with a considerably higher sampling resolution than the previous opacity tables provided by \citet{alexander94}. In this work we use opacities for solar abundance ($X = 0.7$, $Z = 0.02$) based on \citet{caffau11} in the temperature range from $500\,K$ to $30000 \,K$.

\begin{figure}
    \centering
         \resizebox{\hsize}{!}{\includegraphics{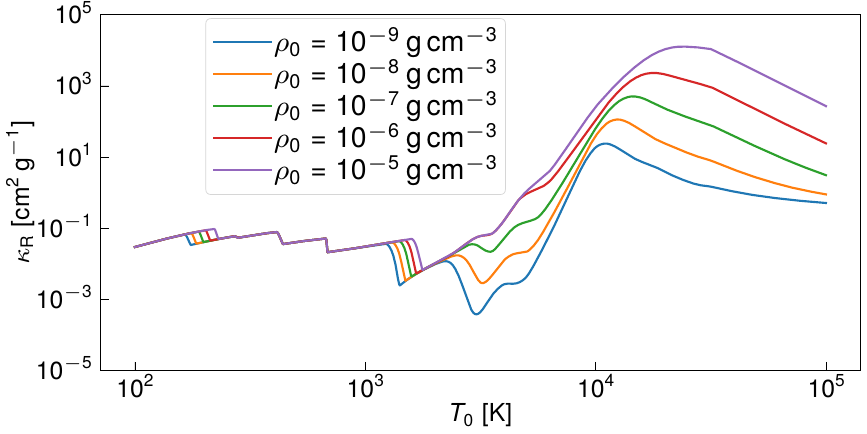}}
    \caption{Rosseland-mean opacities $\kappa_\mathrm{R}$ for various midplane gas densities $\rho_0$.}
    \label{fig:opacities}
\end{figure}

Opacities below $500 \,K$ are dust-dominated, and therefore we use the dust opacity model introduced by \citet{pollack85} for very low temperatures. Dust grains are believed to consist of silicates, iron, troilite, organics, and ice \citep[see e.g.,][]{pollack94, henning96}, which can have different metal abundances and shapes. \citet{pollack85} divided silicates into iron-poor, iron-rich, and normal  species, which have Fe/(Fe + Mg) ratios of 0.0, 0.4, and 0.3, respectively. These latter authors also considered different shapes as spherical and aggregate grains, as well as a distinction between homogeneous, composite, and porous dust grains. For our calculations, the dust-to-gas ratio $f_\mathrm{dust}$ is kept constant in the whole disk and in time, which is an oversimplification. However, we do not treat gas and dust separately in our simulations and cannot calculate a time-dependent and radially changing $f_\mathrm{dust}$. The total Rosseland-mean opacity $\kappa_\mathrm{R}$ for various midplane gas densities $\rho_0(r)$ is shown in \fig{fig:opacities} and can be calculated by combining gas and dust opacities,
\begin{align}
    \kappa_\mathrm{R} = \kappa_\mathrm{R,gas} + f_\mathrm{dust}\, \kappa_\mathrm{R, dust} \,.
\end{align}

\section{Method description}
\label{sec:method_description}
The reasons for using a 1D implicit integration method in this work are briefly explained in \sref{sec:1D_approach} and the method is described in \sref{sec:implicit_integration}. Some vital details of our numerical method are briefly presented in \sref{sec:numerical_method_details}, whereas the boundary conditions used in our simulation are presented in \sref{sec:numerical_boundary_conditions}. An implicit numerical integration scheme needs an initial model to start the simulation with. The construction of such a model is briefly outlined in \sref{sec:stationary_initial_model}.
\subsection{Justification of the 1D approach}
\label{sec:1D_approach}
In recent years, the advent of 2D and 3D models of protoplanetary disks \citep[e.g.,][]{zhu09b, vorobyov07, Vorobyov17, kadam19, zhu19} has lead to a detailed understanding of the complicated interactions of equally important physical processes abundant in such star--disk systems. Hence, the question must be asked, what new insights can be added by another 1D approach.

The study of magnetically truncated, self-consistently simulated protoplanetary disks cannot (at the time of writing) be combined with an investigation of the influence of magnetic fields on the long-term disk behavior. This is because of severe time-step limitations for simulations of the inner disk. An implicit integration scheme (see \sref{sec:implicit_integration}) is not limited to those restrictions (see~Appendix~\ref{sec:imp_timestep}).

We understand that global protoplanetary disk simulations outside the very inner regions ($r > 0.5$ AU) should be at least 2D for studies of the gravitationally unstable phases of disk evolution \citep[e.g.,][]{Vorobyov19}, and 3D for a detailed understanding of complex features such as turbulence for example \citep[e.g.,][]{flock16}. However, in the region within approximately $0.5$ AU,  those models are confined to coarse simplifications (e.g., to the viscous diffusion ansatz or an inner "smart" sink-cell approach) and cannot include important features such as the effects of torques caused by a protostellar magnetic field.
Besides time-step limitations imposed by the CFL condition, multi-dimensional models are either confined to a limited, specific simulation region compared to the whole disk dimensions \citep[e.g.,][]{Vorobyov19} or tend to evolve towards a quasi-stationary state after several orbital periods of the inner disk \citep[e.g.,][]{Romanova04, zhu19}. Both cases are not ideal for investigating the long-term effects over several million years in the innermost disk region.
Additionally, the increased shearing forces towards smaller radii tend to smooth out instabilities in angular directions \citep[see e.g.,][]{Lesur2020} in the very inner disk.

Using an implicit numerical method allows consistent treatment of the inner disk by including the effects of a stellar magnetic field while simultaneously maintaining the ability to carry out long-term evolution studies. Additionally, the self-consistent treatment of large-scale magnetic fields during protoplanetary disk evolution as well as the interaction of the stellar magnetic field and the inner disk makes it necessary to solve the toroidal velocity component $u_\mathrm{\varphi}$ in a time-dependent manner in order to cover magnetic torques \citep[see e.g.,][]{Lubow1994, Guilet2012, Guilet2014}. Disk winds are not included in our simulations but will be addressed separately in a follow-up paper. Apart from a more realistic handling of the radial gas flow, this further supports the choice of solving all hydrodynamic quantities for long-term evolution studies of protoplanetary accretion disks.

The hydrodynamic equations for protoplanetary disks are formulated as a boundary value problem. It is equally true for explicit and implicit methods that boundary conditions for the column density $\Sigma$, the gas velocity $\vec u,$ and the internal energy $e$ imposed on the disk at the inner and outer boundaries determine the radial structure and influence its evolution. An implicit time integration on the other hand will not converge towards a solution at a new time, if the boundary conditions are chosen such that no physical radial structure for $\Sigma$, $\vec u,$ and $e$ can be found for the equations involved. Therefore, an implicit method, despite being more difficult to set up, provides feedback with respect to the physical correctness of the boundary conditions used \citep[see][for further details]{Dorfi1987, stoekl14a}.

Consequently, a 1D implicitly integrated simulation of the inner disk can indeed lead to a better understanding of important features and could for example be used as the inner boundary for more sophisticated 2D and 3D models by providing a physically more accurate description of the very inner regions \citep[e.g.,][]{crida07}.

\subsection{Implicit integration scheme}
\label{sec:implicit_integration}
A numerical method utilizing an implicit integration scheme \citep[e.g.,][]{Dorfi1987, stoekl14a} requires a substantial amount of additional work in development compared to explicit methods. This is because, for each iteration in time $\Delta t$, a Jacobian has to be constructed, which consists of the derivations of each variable at a certain grid point with respect to each variable at every grid point in the computational domain. This matrix must then be subsequently inverted as often as a multidimensional Newton-Raphson iterator needs to converge towards a new solution of the problem at a new time $t + \Delta t$. Another complication is the requirement of an initial model that already solves the discretized set of algebraic equations. The additional work for an implicit integration scheme has the advantage of not being time-step-limited by the Courant-Friedrichs-Levy (CFL) condition. This translates into a condition that information is not transported over more than one grid cell during a single time-step.
This constraint effectively limits the time-step of every hydrodynamic disk simulation with an explicit time-integration scheme. 
Comparing the time-step of our implicit method to the constraint dictated by the CFl condition (see Appendix~\ref{sec:imp_timestep}), we can confirm that the simulation time would be increased by several orders of magnitude to achieve the same radial resolution (2000 radial points are used throughout the simulations in this paper) in the inner regions of the disk with an explicit method.
Consequently, full-blown 3D MHD calculations are very time-consuming and can only be conducted for between a few hundred and a maximum of one thousand orbital periods \citep[see e.g.,][]{romanova14, zhu19}, which would result in an inner disk radius of $r_\mathrm{in} = 0.06$~AU for a covered simulation time-period of $t_\mathrm{total} \leq 15$~years.

\subsection{Numerical method details}
\label{sec:numerical_method_details}
The equations are formulated in one-dimension in the radial direction, are axisymmetric and conservative, and are integrated using a finite-volume method. Also,  a van Leer advection scheme \citep{vanleer77} is used at the cell boundaries. This ensures that the method is second order in space, while the same accuracy in time is achieved by using time-centered variables \citep[for details see e.g.,][]{Dorfi1998, Dorfi2006}.

The equations are discretized by using a staggered mesh \citep[see][]{ragossnig20}, where scalar entities as column density $\Sigma$ and internal energy $e$ are discretized on a scalar mesh while vector-like quantities as the gas velocity component $u_\mathrm{r}$ are discretized on the vector mesh, which is spatially displaced from the scalar mesh by half a grid cell. The velocity component $u_\mathrm{\varphi}$ is of vectorial nature, because for the degeneration of grid cell faces in the angular direction for a 1D radial treatment, $u_\mathrm{\varphi}$ has to be discretized at the scalar mesh. In addition to the numerical treatment of the physical equations, we use an adaptive numerical grid. We apply the grid point distribution introduced by \citet{Dorfi1998}, which avoids the need for coping with artificially introduced perturbations due to grid adaptations, because the mesh configuration is simultaneously solved with the equations. The specific details of our numerical method as well as the detailed discretizations performed for the physical set of equations presented in equations \equs{eq:cont}{eq:ene} and the implementation of the grid equation can be studied in more detail in \citet{ragossnig20}.
\subsection{Numerical boundary conditions}
\label{sec:numerical_boundary_conditions}
Protoplanetary disk quantities such as surface density structure $\Sigma(r)$, disk mass $M_\mathrm{disk}$ , and accretion flow $\dot M(r)$ sensitively depend on the choice of the inner and outer boundary conditions imposed on the disk. At the inner boundary, magnetic torques caused by the stellar magnetic field brake the disk in the angular direction, and consequently the radial flow starts to accelerate towards the star. At a certain point, the magnetic field dominates the disk dynamics and the accretion transitions from a radial drift into an accretion stream along magnetic funnels. The simulation of this magnetically dominated region is intrinsically 3D and cannot be tackled by a 1D approach \citep[see e.g.,][]{bouvier07, romanova14}.
\citet{hartmann16} argued that the disk approximation fails at the magnetic truncation radius $r_\mathrm{trunc}$ and is therefore chosen as our inner boundary $r_\mathrm{in}$ as defined in \equ{eq:truncation_radius}.

A self-consistent treatment of the inner boundary requires the angular velocity to adapt to the magnetic torque of the stellar magnetic field. This is best represented with a zero gradient condition,
\begin{align}
    &\left.\pdv{u_\mathrm{\varphi}}{r} \right|_{r = r_\mathrm{in}} = 0 \; , \label{eq:ibc_uphi}  \\
    &\left.\pdv{u_\mathrm{r}}{r} \right|_{r = r_\mathrm{in}} = 0 \; \label{eq:ibc_ur} .
\end{align}

The gas temperature structure at the inner boundary is chosen such that the sum of contributions of viscous heating and stellar irradiation determines the gas temperature at the inner boundary, which is mapped by employing a Van Neumann boundary condition for the midplane gas temperature $T_0$ at the inner boundary,
\begin{align}
    \left. \pdv{T_0}{r} \right|_{r = r_\mathrm{in}} &= 0 \; .
\end{align}
At the inner boundary, the surface density $\Sigma_\mathrm{in}$ must be able to adapt to different mass transport rates  $\dot M_\mathrm{in}$, for example during an episodic accretion event. Hence, we choose a Van Neumann boundary condition for $\Sigma_\mathrm{in}$,
\begin{align}
    \boldsymbol{\left. \pdv{\Sigma}{r} \right|_{r = r_\mathrm{in}}} &\boldsymbol{= 0 \; .}
\end{align}
At the outer boundary, a mass flux $\dot M_\mathrm{out}$ is assumed to act as an external mass reservoir for the disk.

\begin{align}
    \left. \pdv{\Sigma}{r} \right|_{r = r_\mathrm{out}} &= 0 \;, \\
    u_\mathrm{r}(r_\mathrm{out}) &= \frac{\dot M_\mathrm{out}}{2 \, \pi \, r_\mathrm{out} \, \Sigma(r_\mathrm{out})} \;.
\end{align}
The angular velocity $u_\mathrm{\varphi}$ is assumed to be Keplerian and a Van Neumann boundary condition is used for the internal energy $e$ at the outer boundary condition, which results in
\begin{align}
    \left. \pdv{e}{r} \right|_{r = r_\mathrm{out}} &= 0 \;, \\
    u_\mathrm{\varphi}(r_\mathrm{out}) &= \Omega_\mathrm{K}(r_\mathrm{out}) \, r_\mathrm{out} \;,
\end{align}
where $\Omega_\mathrm{K}(r_\mathrm{out})$ is the Keplerian velocity at the outer boundary. In this study, we focus on the star--disk interaction, and therefore an outer radius of 30~AU is chosen for all simulations.

\subsection{Stationary initial model}
\label{sec:stationary_initial_model}
\begin{figure}[ht]
    \centering
         \resizebox{\hsize}{!}{\includegraphics{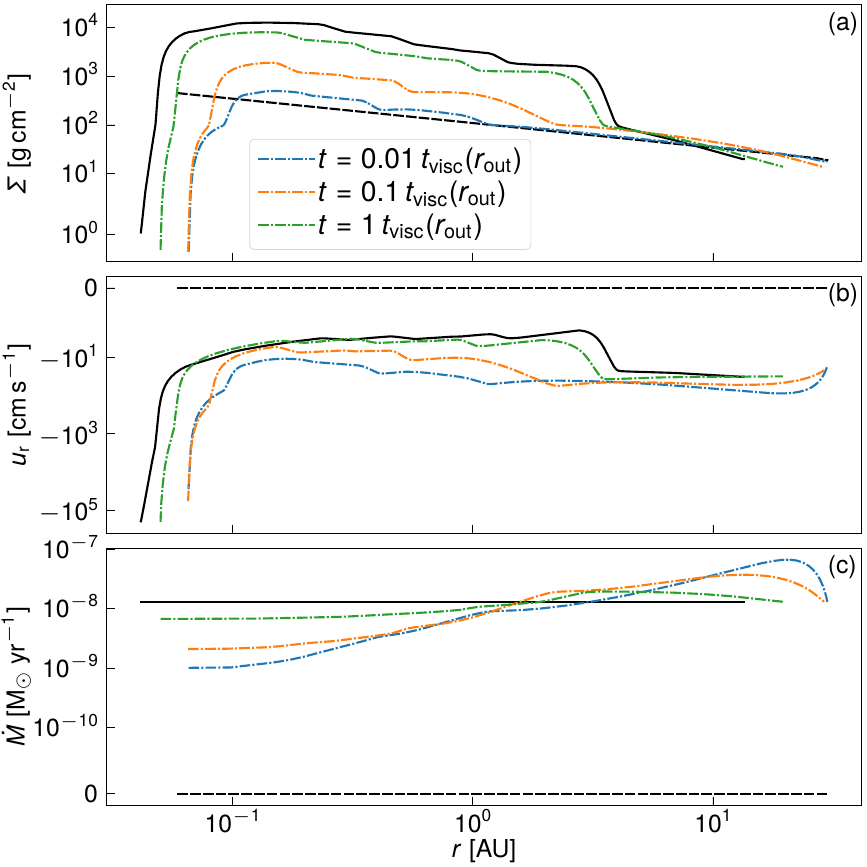}}
    \caption{Snapshots in time, which show the convergence towards a stationary initial model for (a) the surface density $\Sigma(r)$, (b) the radial velocity $u_\mathrm{r}(r),$ and (c) the mass transport rate $\dot M$. The black dashed line shows the initial Keplerian disk, and the black solid line denotes the final stationary model.}
    \label{fig:fiducial_dynamical_relax}
\end{figure}

An implicit numerical method requires an initial model to start with, which already solves the discretized equations along with the grid equation described by \citet{Dorfi1998}. Although, in principle, every solution to the set of equations can be used as a starting model, we aim to construct stationary solutions to start our simulations. Such solutions can be used for verification of the numerical method, because analytical results exist for the assumption of time-independence \citep[see e.g.,][]{armitage10}.

We construct these stationary solutions in a two-stage process. At first, we choose a Keplerian disk with no radial velocity $u_\mathrm{r}$ (see~\fig{fig:fiducial_dynamical_relax}). This disk is in local thermal equilibrium (LTE) in the vertical and radial directions. Second, we introduce a time-independent mass flow across the outer boundary $\dot M_\mathrm{out}$. Collectively, a certain outer mass flux, other boundary conditions, stellar parameters (mass $M_\star$, luminosity $L_\star$, magnetic field $B_\star$, rotation period $P,$ and radius $R_\star$), and viscosity model parameters (cp.~\sref{sec:visco_model}) fully determine the internal radial structure of the disk.

The transition from the starting model to the stationary model occurs on the viscous timescale $t_\mathrm{visc}$,
\begin{align}
    t_\mathrm{visc}(r) &= \frac{r^2}{\nu(r)} \,.
\end{align}
In \fig{fig:fiducial_dynamical_relax} this transition can be seen as mass is transported through the disk, fed by $\dot M_\mathrm{out}$. After $t \gtrapprox t_\mathrm{visc}(r_\mathrm{out}),$ the disk becomes stationary and has adjusted to the boundary conditions and the mass transport rate of the stationary initial model $\dot M_\mathrm{init}$ is constant throughout the whole disk, and therefore $\dot M_\mathrm{init}$ is equal to $\dot M_\mathrm{out}$ (cp.~panel~(c)~of~\fig{fig:fiducial_dynamical_relax}). We also note that $r_\mathrm{in}$ adjusts accordingly for a changing value of $\dot M$ (cp.~\fig{fig:fiducial_dynamical_relax} and \sref{sec:position_IBC}).

We want our simulations to trigger thermal instability and to develop an episodic accretion onto the protostar. During construction of the initial model, this behavior is undesired, and therefore outbursts have to be prevented. We achieve this by increasing the activation temperature $T_\mathrm{active}$ during the first stage to a value that the disk gas temperature $T_0$ does not surpass at any radius. As the model becomes stationary, the time-step can rise due to the implicit nature of the numerical method. Contrary to explicit numerical methods, time-steps can become larger than the viscous timescale $t_\mathrm{visc}$, because they are not restricted by the CFL condition. This effectively subdues the thermal instability, which develops on a much shorter thermal timescale. Therefore, after $\Delta t > t_\mathrm{visc}(r_\mathrm{out})$, the initial model construction enters the second stage, where we smoothly decrease $T_\mathrm{active}$ to the desired value and let the disk model become stationary. We note that this procedure is not physical but is utilized to generate a quasi-stationary initial model, which solves the equations of hydrodynamics.

\section{The importance of the inner disk region}
\label{sec:inner_boundary}

The limitations imposed on explicit numerical methods by the CFL condition (see \sref{sec:numerical_method_details}) are mitigated in \citet{bell94} and subsequent works \citep[e.g.,][]{armitage01, zhu09} by using a diffusion ansatz with a low spatial resolution (compared to our approach) for describing the radial disk evolution. In \sref{sec:revisiting_diffusion_equ} the differences between a full hydrodynamic simulation and the diffusion ansatz are investigated. The importance of the pressure force is evaluated in \sref{sec:press_gradient} and the position of the inner disk boundary is discussed in \sref{sec:position_IBC}.

\subsection{Revisiting the diffusion equation}
\label{sec:revisiting_diffusion_equ}

The radial evolution equation reads \citep[e.g.,][]{pringle81},

\begin{align}
    \pdv{\Sigma}{t} &= -\frac{1}{r} \pder{r} \left[ \frac{1}{(r^2 \, \Omega)'} \pder{r} (\nu \, \Sigma \, r^3 \, \Omega') \right] \;, \label{eq:diff_eq}
\end{align}

where $\Omega(r)$ corresponds to the angular velocity. In the derivation of \equ{eq:diff_eq}, pressure gradients are not included in the angular momentum equation \citep[e.g.,][]{pringle81}. If $\Omega(r) \propto \Omega_\mathrm{K}(r)$, with $\Omega_\mathrm{K}(r)$ depicting the Keplerian velocity, \equ{eq:diff_eq} becomes

\begin{alignat}{2}
    &\pdv{\Sigma}{t} &&= \frac{3}{r} \pder{r} \left[ \sqrt{r} \, \pder{r} (\nu \, \Sigma \, \sqrt{r} ) \right] \;, \label{eq:diff_eq_simple} \\
    &u_\mathrm{r} &&= -\frac{3}{ \Sigma \, \sqrt{r}} \, \pder{r} (\nu \, \Sigma \, \sqrt{r} ) \;, \label{eq:diff_eq_simple_ur}
\end{alignat}

where the continuity equation \equ{eq:cont} is used to obtain a formulation for the radial drift velocity $u_\mathrm{r}$. Torques due to for example protostellar magnetic fields cannot be treated in a self-consistent way with \equ{eq:diff_eq_simple}, because this would involve braking (or acceleration) of the angular velocity component in a way in which $\Omega(r)$ is no longer proportional to $\Omega_\mathrm{K}(r)$ . In order to see the impact of a perturbation in the angular velocity profile $\Omega(r)$, one can model the angular velocity as Keplerian with a small linear perturbation $\Omega_\mathrm{1}$,

\begin{align}
   \Omega(r) = \Omega_\mathrm{K}(r) + \Omega_\mathrm{1}(r) \;.
\end{align}

After linearization of \equ{eq:diff_eq} with respect to $\Omega_\mathrm{1}$, the following expression is obtained:

\begin{alignat}{2}
    &f_\mathrm{corr} &&= \frac{2 \, ( r^2 \Omega_\mathrm{1} )' }{r \, \Omega_\mathrm{K}} + \frac{2 \, ( \nu \, \Sigma \, r^3 \Omega_\mathrm{1}' )' }{3 \, ( \nu \, \Sigma \, r^2 \Omega_\mathrm{K} )'} \;,  \label{eq:diff_eq_fcorr} \\
    &\pdv{\Sigma}{t} &&= \frac{3}{r} \pder{r} \left[ \sqrt{r} \, \pder{r} (\nu \, \Sigma \, \sqrt{r} ) \left( 1 -  f_\mathrm{corr} \right) \right] \;. \label{eq:diff_eq_linearization} \\
    &u_\mathrm{r} &&= -\frac{3}{ \Sigma \, \sqrt{r}} \, \pder{r} (\nu \, \Sigma \, \sqrt{r} ) ( 1 - f_\mathrm{corr} ) \;, \label{eq:ur_eq_linearization}
\end{alignat}
where a primed quantity $A'$ denotes its radial derivative $\partial A/\partial{r}$. The first term of the correction factor $f_\mathrm{corr}$ depends on the radial gradient of $\Omega_\mathrm{1}$; hence a radially localized, sharp perturbation in $\Omega(r)$ can lead to a non-negligible effect. During the onset of an FU Ori-like outburst, an ionization front is propagating radially inwards and outwards through the disk \citep[for a detailed discussion see][]{bell94}. Along these waves, the disk is adapting itself to a changed disk temperature and viscosity. Various authors \citep[e.g.,][]{bell94, armitage10, zhu09} have shown that those waves occur in the column density $\Sigma(r)$, the internal energy profile $e(r),$ and in the radial mass transport rate $\dot M(r)$. However, the ionization front is also visible as a wave in the radial angular velocity profile $\Omega(r)$, which then modifies the accretion rate and subsequently the density profile. The second term in \equ{eq:diff_eq_fcorr} takes into account that a modified relative angular velocity between two neighboring radial disk annuli also changes the shear that those disk rings are experiencing. $f_\mathrm{corr}$ can be interpreted as the deviation from the viscous contribution to the diffusion equation. In \equ{eq:ur_eq_linearization} a correction factor $f_\mathrm{corr} > 1$ leads to a radially localized, outward-bound mass flow, which is caused by redistribution of angular momentum in sharp angular velocity perturbations $\Omega_1(r)$.

\begin{figure}[ht]
    \centering
         \resizebox{\hsize}{!}{\includegraphics{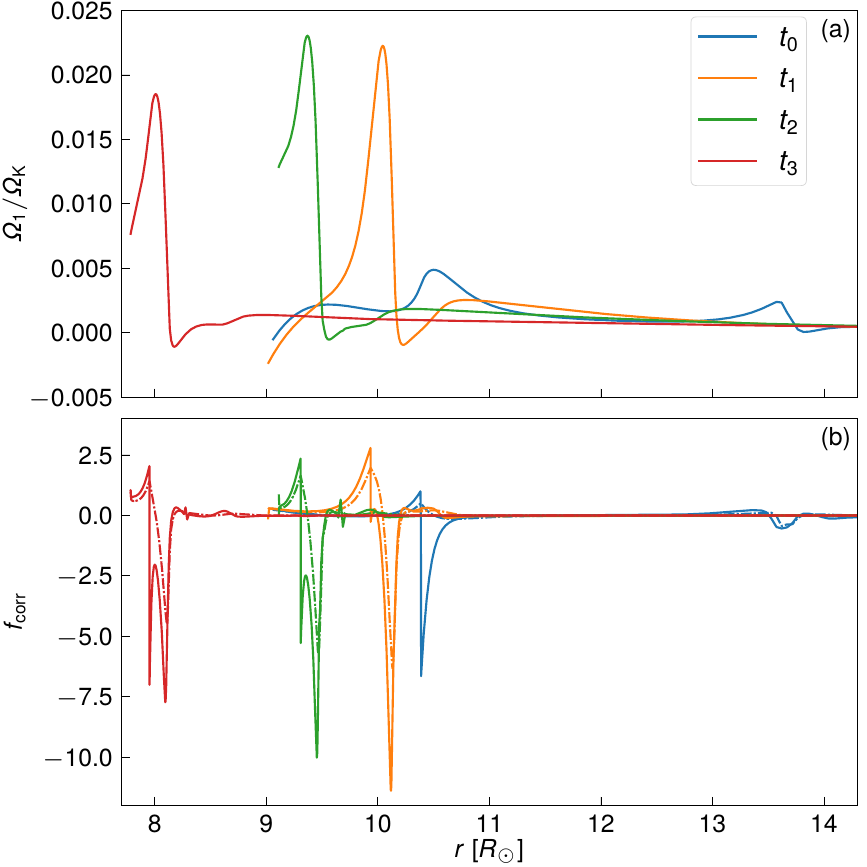}}
    \caption{(a) $\Omega_\mathrm{1}$ waves for four different, arbitrarily chosen times $t_0$ to $t_3$ during the onset of a FU Ori-like outburst. (b) Correction factors as defined in \equ{eq:diff_eq_fcorr} for the same times as in (a). The dashed-dotted lines correspond to the first term in \equ{eq:diff_eq_fcorr}, whereas the full lines show the total value of $f_\mathrm{corr}$.}
    \label{fig:fiducial_burst_fcorr}
\end{figure}

In \sref{sec:episodic_accretion} we provide a detailed description of the perturbations during the onset of a FU Ori-like outburst, whose influence on $f_\mathrm{corr}$ is shown in \fig{fig:fiducial_burst_fcorr}. An angular velocity perturbation $\Omega_\mathrm{1}$ induced by sharp peaks of MRI efficiency due to density waves causes a radially localized peak in $|f_\mathrm{corr}|$. Four snapshots are taken for four different times $t_0 < t_1 < t_2 < t_3$ immediately after the onset of TI and shows the inward-bound ionization front (cp.~\sref{sec:episodic_accretion}). The first term of \equ{eq:diff_eq_fcorr} is plotted separately in \fig{fig:fiducial_burst_fcorr} to show the direct influence of a nonKeplerian velocity $u_\mathrm{\varphi}$ compared to the additional influence of altered shear between two neighboring disk annuli. For a yet unpronounced perturbation at time $t_0$, the modified shear clearly dominates $f_\mathrm{corr}$, but at later times $t_1$ to $t_3$ the direct influence of $\Omega_1$ is also non-negligible. While the diffusion equation approach \equ{eq:diff_eq_simple} appears to be well suited to describing the disk evolution in the absence of such waves, without solving the momentum equation in angular direction, there are, at least locally, substantial deviations to be expected. 

\begin{figure}[ht]
    \centering
         \resizebox{\hsize}{!}{\includegraphics{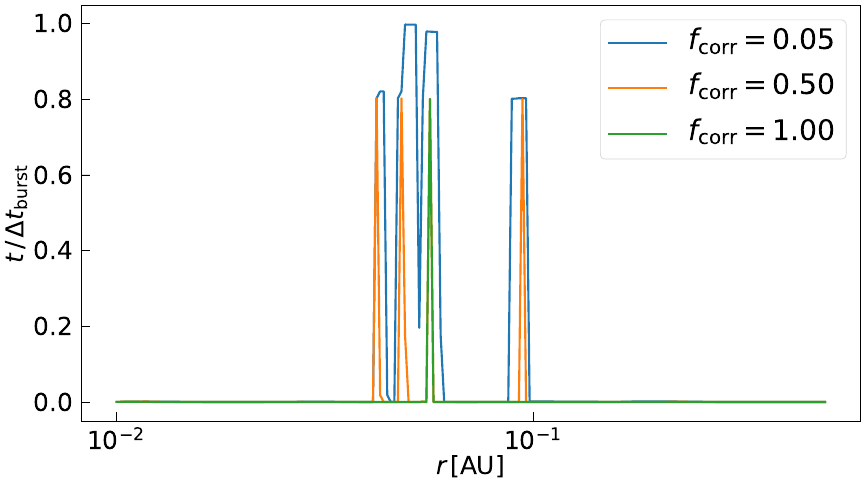}}
    \caption{Time relative to the burst duration in the fiducial model $\Delta t_\mathrm{burst}$, in which a region in the inner 0.5~AU exceeds a certain $f_\mathrm{corr}$ threshold (0.05, 0.5 and 1.0).  The radial range is divided in 200  equally spaced logarithmic bins.}
    \label{fig:fcorr_time}
\end{figure}

The deviations in \fig{fig:fiducial_burst_fcorr} represent a snapshot during time evolution at a certain time $t$, which can affect the short term behaviour of the disk. A more qualitative way to investigate how strongly a certain disk region is affected by $f_\mathrm{corr}$ over a longer period of time is shown in \fig{fig:fcorr_time}. During the duration of an outburst ($\Delta t_\mathrm{burst} \sim 20$ years) in our fiducial model, we have added up the times in which the absolute value of $f_\mathrm{corr}$ is larger than a certain threshold (0.05, 0.5 and 1.0). In the innermost disk ($\lesssim 0.1$~AU) $f_\mathrm{corr}$ can exceed 0.05 over 90\% of $\Delta t_\mathrm{burst}$. Even values of $f_\mathrm{corr} = 0.50$ and $f_\mathrm{corr} = 1.00$ are exceeded during 80 \% of $\Delta t_\mathrm{burst}$ in some regions of the inner disk.

\subsection{Importance of the pressure gradient}
\label{sec:press_gradient}

\begin{figure}[ht]
    \centering
         \resizebox{\hsize}{!}{\includegraphics{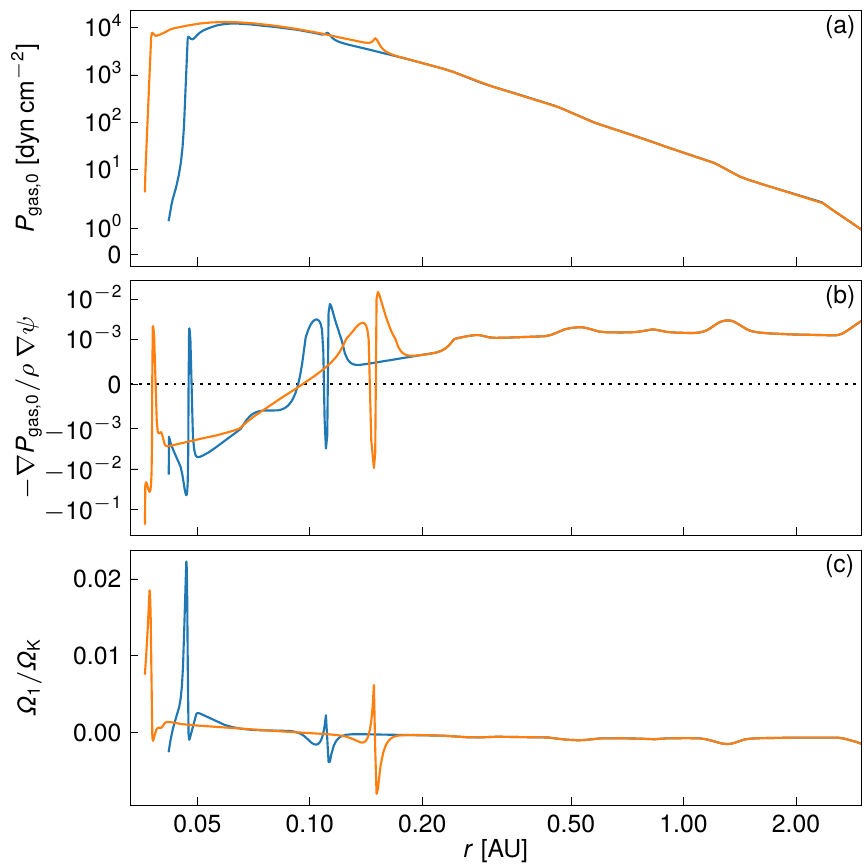}}
    \caption{(a) Midplane gas pressure $P_\mathrm{gas,0}$ between $0.04$ AU and $2$ AU for two different times during an FU Ori like outburst. (b) The midplane pressure gradient force $-\nabla P_\mathrm{gas,0}$ relative to the gravitational force is plotted. The radially localized changes in sign occur at the positions of the ionization fronts running inwards and outwards. The deviation from the Keplerian angular velocity is shown in (c). } 
    \label{fig:fiducial_burst_pgas}
\end{figure}

In a steady-state protoplanetary disk the gas rotation velocity $u_\varphi$ (neglecting radial viscous forces, magnetic fields, and radial advection of momentum) is essentially given by
\begin{align}
    \frac{u_\varphi^2}{r} &\approx \frac{G \, M_\star}{r^2} + \frac{1}{\rho} \frac{dP}{dr} \label{eq:approx_eqmot_steady_state} \;.
\end{align}
Therefore, in order to have a subKeplerian steady-state accretion disk, the pressure gradient has to yield $dP / dr < 0$. For all radii but the very innermost parts ($r < 0.1$ AU), this condition is usually fulfilled. Furthermore, the pressure gradients are of minor order compared to the centrifugal forces, and at the inner boundary a zero-torque boundary condition is usually applied for such models in the absence of a stellar magnetic field. For example the IBC applied by \citet[][]{bell94},
\begin{align}
    \Sigma(r = r_\mathrm{in}) \approx 0 \;,
\end{align}
models a disk extending to the stellar surface and therefore accounts for the transition of a protoplanetary disk to $u_\mathrm{r} = 0$. However, a magnetically braked inner disk has an increased accretion rate and therefore a lower column density $\Sigma$ and diminished gas pressure $P_\mathrm{gas,0}$ close to the inner rim (see \fig{fig:fiducial_burst_pgas}). Hence, in the very innermost regions ($r \lessapprox 0.2$ AU) the pressure gradient force $-\gradient P$ (cp.~\equ{eq:mot}) helps the gravitational force to push material inwards, which leads to a super-Keplerian $u_\mathrm{\varphi}$ according to \equ{eq:approx_eqmot_steady_state}. This can also be seen in \fig{fig:fiducial_burst_fcorr}, where $\Omega_\mathrm{1} > 0$ (except for the declining slopes of the ionization fronts). The pressure gradient force also exceeds the viscous force contribution in the region of the ionization front (see \fig{fig:fiducial_burst_forces} in \sref{sec:episodic_accretion}).
In the very inner regions, the diffusion equation ansatz of \equ{eq:diff_eq_simple} is less suitable for describing the time evolution of the disk. This is especially true for thermally unstable disks (cp.~\fig{fig:fiducial_burst_fcorr} and \fig{fig:fiducial_burst_pgas}), where due to localized perturbations of $u_\mathrm{\varphi}$ the deviations from the diffusion equation \equ{eq:diff_eq_simple} are non-negligible (see \fig{fig:fcorr_time}).

\subsection{The position of the inner disk boundary}
\label{sec:position_IBC}

The inner disk is coupled to the star via the stellar magnetic field, which means that, inside the corotation radius $r_\mathrm{cor}$, the angular momentum of the disk is transferred to the protostar, whereas outside of $r_\mathrm{cor}$ the disk is accelerated. This star--disk connection results in mutual transfer of angular momentum between the star and the inner disk \citep[e.g.,][]{matt2010}. However, in the present study,  the spin-up or spin-down of the  star due to gain or loss of angular momentum  is not considered, as our focus is on the influence of stellar magnetic torque on the inner disk. The inclusion of angular momentum transfer introduces a further complexity which makes it more difficult to separate the various influences altering the long-term evolution of a protoplanetary disk and has to be investigated in further studies.

\begin{figure}[ht]
    \centering
         \resizebox{\hsize}{!}{\includegraphics{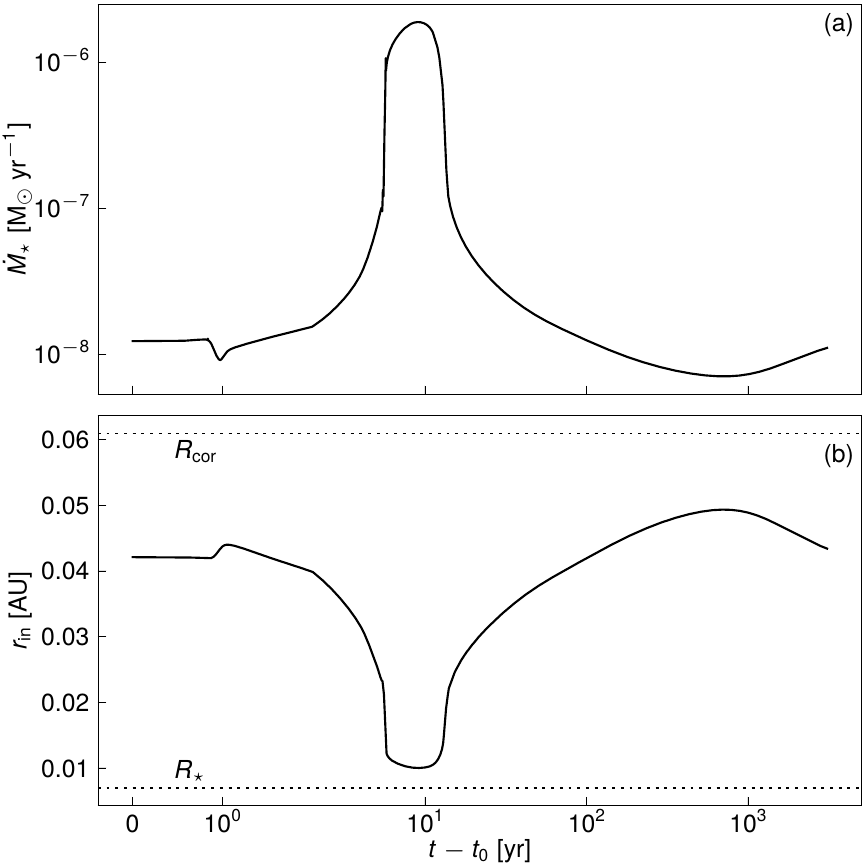}}
    \caption{(a) Accretion rate for a full TI-triggered outburst for our fiducial model \texttt{fid} (cp.~\tab{tab:model_parameters}). (b) Variation of the inner boundary during the burst is shown in (a). Here, $t_\mathrm{0}$ corresponds to the onset of the first burst. The two dashed lines denote the protostellar radius $R_\star$ and the corotation radius $r_\mathrm{cor}$.} 
    \label{fig:fiducial_burst_rin}
\end{figure}

The corotation radius is defined according to Kepler's third law, 
\begin{equation}
    r_\mathrm{cor} = \left( \frac{GM_\star P^2}{4 \pi^2} \right)^{1/3} \label{eq:corotation_radius} \;,
\end{equation}
where $P$ is the stellar rotational period. Following \citet[][]{herbst01}, we choose the stellar rotation period  $P = 6$~days such that $r_\mathrm{cor} \approx 0.06$~AU for all simulations in this work. 
Magnetic torques tend to slow down the disk inside $r_\mathrm{cor}$ and the disk's toroidal velocity $u_\mathrm{\varphi}$ becomes subKeplerian. Consequently, the disk material inside $r_\mathrm{cor}$ starts to radially accelerate towards the star until it reaches the magnetic truncation radius $r_\mathrm{trunc}$. Inside $r_\mathrm{trunc}$, the magnetic pressure of the stellar dipole field exceeds the ram pressure $P_\mathrm{ram}$ of the infalling material, and therefore the disk gets disrupted and accretes along magnetic funnels onto the star. Consequently, $r_\mathrm{trunc}$ is a natural choice for the inner disk edge in the case of strong accretion, where $P_\mathrm{ram}$ exceeds the gas pressure $P_\mathrm{gas}$. Consequently, $r_\mathrm{in}$  is given as follows \citep[for details cp.][]{hartmann16}:
\begin{align}
    r_{\mathrm{in}}(P_\mathrm{gas} < P_\mathrm{ram}) \approx 18 \, \xi \, R_\odot \, & \left(\frac{B_\star}{10^3 \, G}\right)^{4/7}  \left(\frac{R_*}{2 \, R_\odot}\right)^{12/7} \left(\frac{M_\star}{0.5 \, M_\odot}\right)^{-1/7} \nonumber \\
    & \left(\frac{\dot M_\star}{10^{-8} \, M_\odot / \mathrm{yr}}\right)^{-2/7} \label{eq:truncation_radius} \;,
\end{align}
where $B_\star$, $R_\star$, and $M_\star$ are the stellar magnetic field at the stellar surface, the protostellar radius, and its mass, respectively, and $\xi$ is a correction factor that accounts for the rather complicated details of disk--star interactions and is usually set to $\xi < 1$ according to \citet[][]{hartmann16}. In this work, we choose $\xi = 0.65$. Further, $\dot M_\star$ denotes the accretion flow over the inner boundary onto the star and is in general time-dependent.
As the inner radius $r_\mathrm{in}$ is dependent on the accretion rate $\dot M_\star$, it is likely to move inwards during increased accretion, for example during a FU~Ori-like burst. The magnetic field cannot withstand the high accretion rate during an outburst event, and therefore the stellar magnetosphere gets squashed closer to the star. The movement of the inner boundary according to the time-dependent mass accretion rate $\dot M_\star(t)$ for our fiducial model (see~\tab{tab:model_parameters} and \sref{sec:episodic_accretion}) during an outburst is shown in \fig{fig:fiducial_burst_rin}. During an outburst, $r_\mathrm{in}$ is pushed towards the stellar radius $R_\star$ (cp.~\equ{eq:truncation_radius}). We want to emphasize that our method is able (from a numerical point of view) to cover the case of $r_\mathrm{in} \sim R_\star$ in combination with long-term simulations.

In regions where the ram pressure $P_\mathrm{ram}$ exceeds the gas pressure $P_\mathrm{gas}$, \equ{eq:truncation_radius} describes the movement of $r_\mathrm{in}$ well. If $P_\mathrm{gas}$ is larger than $P_\mathrm{ram}$, the protostellar magnetic field is able to disrupt the disk as soon as the magnetic pressure $P_\mathrm{mag} = B_\mathrm{z}^2 / (8 \, \pi )$ becomes comparable to $P_\mathrm{gas}$, which for a dipole field approximation occurs at
\begin{alignat}{2}
    &r_{\mathrm{in}}(P_\mathrm{gas} > P_\mathrm{ram}) = \frac{{B_\star}^{1/3} \, R_\star}{{P_\mathrm{gas}}^{1/6} \, (8 \pi )^{1/6} } \;. \label{eq:truncation_radius_press}
\end{alignat}
In this work, we find two scenarios in thermally unstable disks where the gas pressure $P_\mathrm{gas}$ exceeds the accretion ram pressure $P_\mathrm{ram}$. The first is after an outburst, when the accretion rate decreases until the disk evolves towards the next burst (cp.~\sref{sec:results}). The second is towards the end of a disk's lifetime when the accretion rate decreases, which yields a lower ram pressure $P_\mathrm{ram}$ (see \sref{sec:influence_stellar_magnetic_field}).

\label{sec:diffusion_equation}

\section{Results and Discussion}
\label{sec:results}

The goal of our simulations is to show the influence of a protostellar magnetic field on the bursting behavior of MRI- and TI-unstable disks. Therefore, in \sref{sec:episodic_accretion} the bursting behavior and the long-term evolution of such disks is investigated. In \sref{sec:influence_stellar_magnetic_field} we perform long-term runs with different magnetic field strengths $B_\star$ to investigate the effect of magnetic torques on the inner disk dynamics and its consequences for the long-term behavior. Additionally, the effect of different MRI activation temperatures is discussed in \sref{sec:influence_tcrit}.

\subsection{Episodic accretion and long-term evolution of magnetically truncated low-mass protoplanetary disks}
\label{sec:episodic_accretion}

\citet{armitage01} and \citet{zhu10a, zhu10b} argue that a protoplanetary disk cannot sustain a steady mass transport rate in the radial direction from $r \approx 100$ AU to its inner edge at a few stellar radii, because GI in the outer disk feeds too much mass to the inner disk where it piles up. This leads to viscous heating in this region of enhanced density and eventually to the activation of the MRI. Enhanced radial mass transport $\dot M$ in the MRI-active disk yields more viscous heating and hence facilitates triggering of the thermal instability \citep[e.g.,][]{bell94}.

\begin{table*}[ht]
\centering
\caption{Simulation run parameters}              
\begin{tabular}{l c c c c c c c c r}         
\hline\hline                       
Model & $M_\star$ [$M_\mathrm{\odot}$] & $R_\star$ [$R_\odot$] & $B_\star$ [kG] & $r_\mathrm{cor}$ [AU] & $M_\mathrm{disk, init}$ [$M_\odot$] & $\dot M_\mathrm{init}$ [$M_\odot \, \mathrm{yr}^{-1}$] & $\alpha_\mathrm{MRI}$ & $\alpha_\mathrm{MRI} / \alpha_\mathrm{base}$ & $T_\mathrm{active}$ [K] \\
\hline                                  
    \texttt{fid} & $1.0$ & $1.5$ & $2.0$ & $0.0609$ &  $10^{-2}$ & $1.3 \cdot 10^{-8}$ & $2.1 \cdot 10^{-2}$ & 50 & 1530 \\
    \texttt{1p5kG} & 1.0 & 1.5 & 1.5 & 0.0609 &  $10^{-2}$ & $1.3 \cdot 10^{-8}$ & $2.1 \cdot 10^{-2}$ & 50 & 1530 \\
    \texttt{3p0kG} & 1.0 & 1.5 & 3.0 & 0.0609 &  $10^{-2}$ & $1.3 \cdot 10^{-8}$ & $2.1 \cdot 10^{-2}$ & 50 & 1530 \\ 
    \texttt{4p0kG} & 1.0 & 1.5 & 4.0 & 0.0609 &  $10^{-2}$ & $1.3 \cdot 10^{-8}$ & $2.1 \cdot 10^{-2}$ & 50 & 1530 \\ 
    \texttt{5p0kG} & 1.0 & 1.5 & 5.0 & 0.0609 &  $10^{-2}$ & $1.3 \cdot 10^{-8}$ & $2.1 \cdot 10^{-2}$ & 50 & 1530 \\ 
\hline                                             
\end{tabular}
\label{tab:model_parameters}  
\end{table*}

In \tab{tab:model_parameters} the parameters for our fiducial model \texttt{fid} are stated. $\dot M_\mathrm{init}$ corresponds to the constant mass transport rate throughout the disk, which is obtained for the stationary initial model constructed as described in \sref{sec:stationary_initial_model}. The protostellar parameters (mass $M_\star$, radius $R_\star$, magnetic dipole field strength $B_\star$, and the corotation radius $r_\mathrm{cor}$), together with the disk parameters ($\alpha$-viscosity parameter due to MRI $\alpha_\mathrm{MRI}$, the ratio $\alpha_\mathrm{MRI} / \alpha_\mathrm{base}$, and the thermal instability activation temperature $T_\mathrm{active}$) and the opacity prescription (cp.~\fig{fig:opacities}), determine the initial disk mass $M_\mathrm{disk,init}$ for the stationary initial model. We choose our parameters such that $M_\mathrm{disk,init} \approx 0.01\,\mathrm{M_\odot}$ to ensure a gravitationally stable disk (see~\sref{sec:theoretical_framework}~and~\sref{sec:visco_model}). For all simulation runs in \tab{tab:model_parameters}, no external mass reservoir is assumed to resemble the configuration of a late class II star--disk system, and therefore $\dot M(r_\mathrm{out}) = 0$.

\begin{figure*}[ht]
    \centering
    \includegraphics[width=18cm]{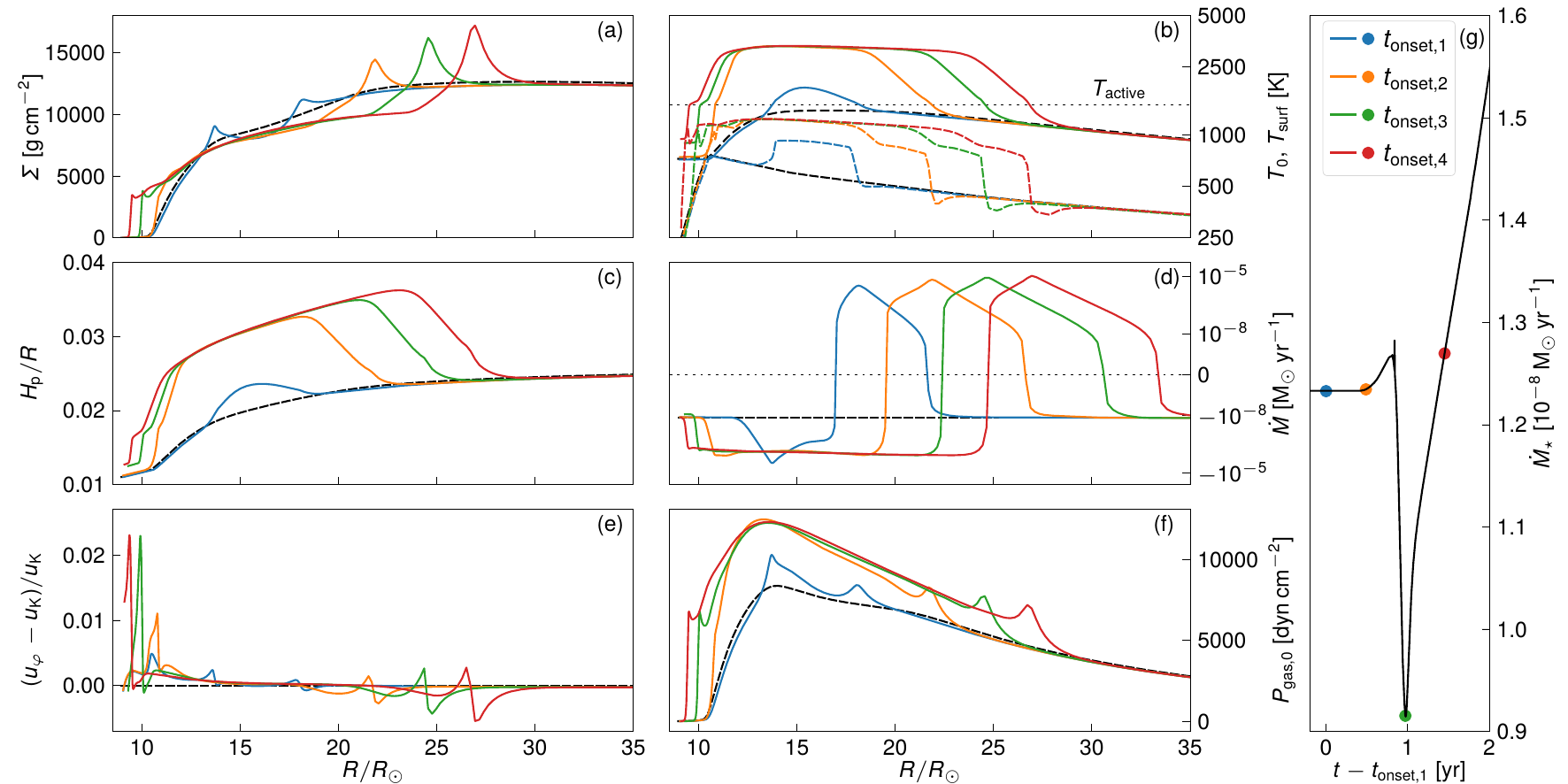}
    \caption{Various disk quantities for certain times $t_\mathrm{onset,1} - t_\mathrm{onset,4}$ describing an outburst onset phase. Panels (a) and (c) show the radial surface density structure $\Sigma$ and the pressure scale-height $H_\mathrm{p}$ during the onset of an outburst, respectively. Panel (b) shows the midplane gas temperature $T_0$ (solid) and the surface temperature $T_\mathrm{surf}$ (dashed). The dotted line marks the activation temperature $T_\mathrm{active}$. In panel  (d) the radial mass transport rate $\dot M$ is shown, where the dotted line corresponds to $\dot M = 0$. The deviation from a Keplerian accretion disk is shown in panel (e), whereas panel  (f) shows the gas pressure at the midplane $P_\mathrm{gas,0}$. Finally, panel (g) maps the various snapshots in time to the stages during the onset of an outburst. The dashed black line in panels (a) to (f) denote the stationary initial model as described in \sref{sec:stationary_initial_model}}.
    \label{fig:burst_fid_onset}
\end{figure*}

The column density $\Sigma$, midplane gas temperature $T_\mathrm{gas}$, scale height $H_\mathrm{p}$, the radial mass transport rate $\dot M$, the relative deviation from the Keplerian velocity $u_\mathrm{K}$, and the midplane gas pressure $P_\mathrm{gas,0}$ of our fiducial model \texttt{fid} are shown in \fig{fig:burst_fid_onset} during onset of a TI-triggered burst. \citet{zhu10a} observed short phases during outbursts in their simulations, where the thermal instability could not be sustained (they termed this phenomenon \textit{drop-outs}), whose cause they interpreted to be the lack of radial advection in their simulations. We include radial advection (cp.~\sref{sec:theoretical_framework}) in our models and do not observe such drop-outs, which confirms their interpretation. Shortly before $t_\mathrm{onset,1}$, a TI is triggered at $r \approx 15\,R_\odot$, which can be seen by a steep rise in $T_\mathrm{gas,0}$ (panel (b) of \fig{fig:burst_fid_onset}). Simultaneously, the TI strongly increases  local viscosity, to which the disk reacts
by redistributing material to both the inner and outer disk annuli \citep[e.g.,][]{bell94}. This increases $\Sigma$ at both neighboring disk annuli above the threshold at which TI is also triggered. Hence, two ionization fronts are starting to travel inwards and outwards, which can be seen most clearly in panels (a), (b), and (e) of \fig{fig:burst_fid_onset} for different points in time from $t_\mathrm{onset,1}$ to $t_\mathrm{onset,4}$. Both ionization fronts heat up the disk (see \fig{fig:burst_fid_onset} (b)) which leads to an increased scale height $H_\mathrm{p}$ (\fig{fig:burst_fid_onset} (c)). This in turn yields a region where the disk is shadowed from protostellar irradiation. Our models incorporate geometric shadowing caused by local elevations in $H_\mathrm{p}$ (cp.~\sref{sec:energy_equation}), which can be seen in \fig{fig:burst_fid_onset} (b) as a drop in surface temperature $T_\mathrm{surf}$ just behind (from the perspective of the  protostar) the outward-traveling ionization front. As soon as the inward-bound wave reaches the  rim of the inner disk, heated (and hence more viscous) material can no longer be redistributed further inwards, and the disk starts to transition into the outburst phase. The gas temperature at the inner boundary $T_\mathrm{gas,0}(r_\mathrm{in})$ rises very quickly and is accompanied by a sharp increase in viscosity and hence in $\dot M_\star$ (cp.~panel (g) of \fig{fig:burst_fid_onset}).

\begin{figure}[ht]
    \centering
    \resizebox{\hsize}{!}{\includegraphics{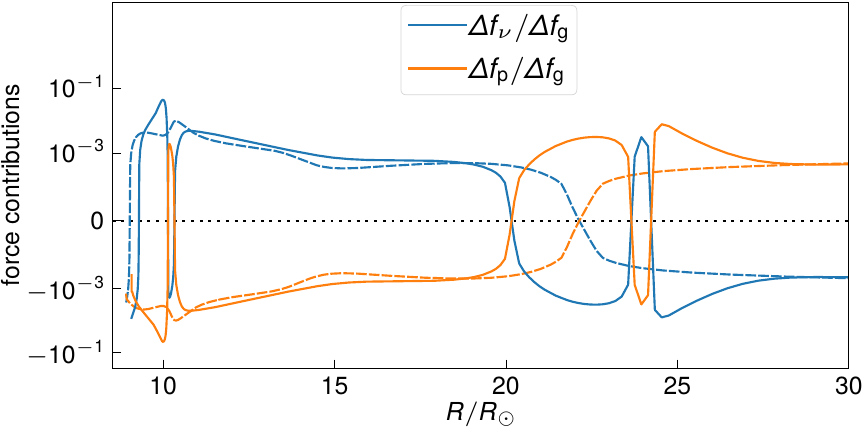}}
    \caption{Net contributions of pressure $\Delta f_\mathrm{p}$ and viscous force $\Delta f_\mathrm{\nu}$ on an disk annulus are shown for the inner disk. The forces are in units of the net gravitational force $\Delta f_\mathrm{g}$ at its corresponding radius. The dashed lines outline the various force contributions for the stationary initial model. The dotted horizontal line shall help to visualize where the force contributions change sign.} 
    \label{fig:fiducial_burst_forces}
\end{figure}

The deviation from the diffusion approach investigated in \sref{sec:revisiting_diffusion_equ} can be seen clearly in panels (e) and (f) of \fig{fig:burst_fid_onset}. The relative difference of $u_\mathrm{\varphi}$ compared to the Keplerian velocity $u_\mathrm{K}$ peaks at $\sim 0.02$, which is an order of magnitude larger than the predicted $u_\mathrm{\varphi} \approx 0.996\, u_\mathrm{K}$ for a viscous protoplanetary accretion disk \citep[see][]{armitage10}. Additionally, the rotational velocity $u_\mathrm{\varphi}$ not only becomes subKeplerian but also superKeplerian (see panel (e) in \fig{fig:burst_fid_onset}). This feature is strongest along the propagating ionization fronts, as there the surface density has localized peaks, which in turn leads to an altered (with respect to a quasiKeplerian accretion disk) viscous net shear on the disk annuli at those radii. The relative influence of gas pressure $P_\mathrm{gas}$ on a certain annulus compared to the viscous force is shown in \fig{fig:fiducial_burst_forces}. The region at around $10$ solar radii in \fig{fig:fiducial_burst_forces} corresponds to the location of the inward-bound ionization front, whereas the outer ionization front is located at approximately $24\, R_\odot$.
The magnetic braking of the inner disk inside $r_\mathrm{cor}$ leads to an increased radial velocity $u_\mathrm{r}$ and hence to a drop in $\Sigma$ towards $r_\mathrm{in}$, which leads to the pressure force changing sign at around $22 \, R_\odot$.
Furthermore, it can be seen in \fig{fig:fiducial_burst_forces} that the gas pressure alters the disk dynamics close to the inner disk radius, which is because that is where the pressure gradient becomes steepest (apart from local disturbances like ionization fronts). The influence of the pressure gradient even exceeds the viscous contributions, which is, once more, an indication that the full set of hydrodynamic equations are required to properly model the inner regions of a disk. The sharp decrease in $\dot M_\star$ between $t_\mathrm{onset,2}$ and $t_\mathrm{onset,3}$ in \fig{fig:burst_fid_onset} occurs because of the small-scale structure of the inbound wave when it reaches $r_\mathrm{in}$ \citep[for a thorough analysis of the thermal instability see][]{bell94}. 

\begin{figure*}[ht]
    \centering
    \includegraphics[width=18cm]{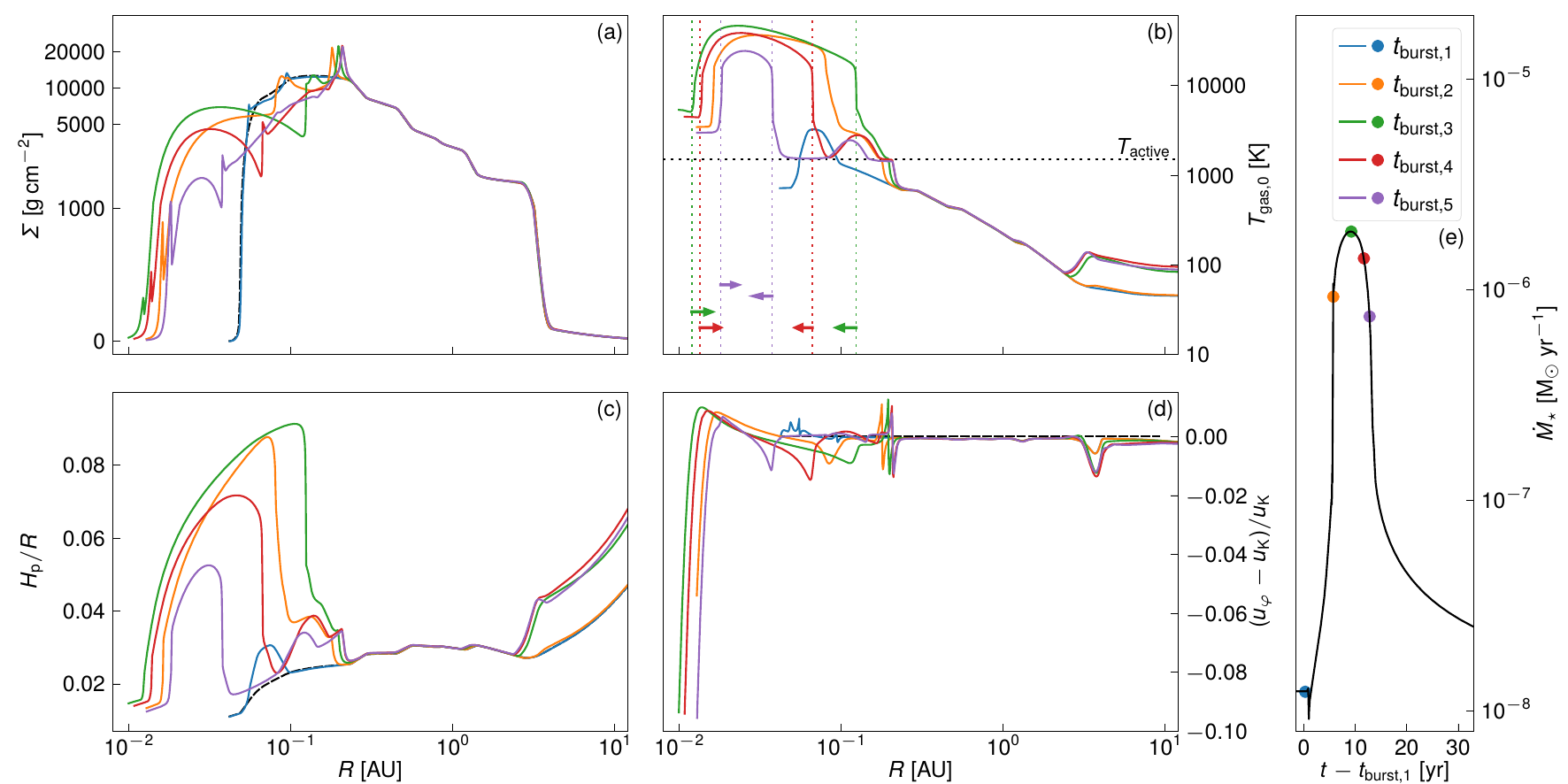}
    \caption{ Panels (a) to (c), as well as (e), are the same disk properties as in \fig{fig:burst_fid_onset}, but for five points in time $t_\mathrm{burst,1}$ to $t_\mathrm{burst,5}$ during an outburst. Panel (d) in this figure shows the same quantity as (e) in \fig{fig:burst_fid_onset}. The dashed black line in panels (a) to (d) denote the stationary initial model as described in \sref{sec:stationary_initial_model}. The radial range in this plot is chosen from $r_\mathrm{in}$ to 10~AU. The color-coded, dashed vertical lines and corresponding arrows mark the position of the cooling waves and their propagation direction, respectively. }
    \label{fig:burst_fid_full}
\end{figure*}

\figl{fig:burst_fid_full} shows the whole disk during a TI-induced outburst. The outward-bound ionization front as described in detail by \citet{bell94} stalls at $t \approx t_\mathrm{burst,4}$ at a radius where the swept-up material is no longer capable of increasing $\Sigma$ sufficiently for the disk annulus to become thermally unstable  (cp.~(b) in \fig{fig:burst_fid_full}). The increased accretion rate yields a higher accretion luminosity, which in turn causes a higher temperature at the disk surface and a puffing-up of the irradiated outer disk (panel (c) of \fig{fig:burst_fid_full}). The detailed effects of the increased scale height, relating to shadowing of the disk behind this bump, will be studied in another paper of this series. The ongoing depletion of the inner disk inside the ionization front due to strongly increased viscosity eventually leads to a drop in gas temperature $T_\mathrm{gas,0}$. At approximately $t_\mathrm{burst,3}$, hydrogen is able to recombine again and a cooling wave starts to propagate inwards (radially outer green vertical dashed line and arrow in panel (b) of \fig{fig:burst_fid_full}). The cooler gas then leads to a drop in optical depth (cp.~\fig{fig:opacities}), which results in termination of the TI. This inward-bound perturbation becomes clearly visible at $t_\mathrm{burst,4}$ and $t_\mathrm{burst,5}$ in panels (a) to (c) of \fig{fig:burst_fid_full}, where the position and direction of the cooling front is marked with a vertical dashed line and an arrow, respectively. As this wave propagates towards the inner disk radius $r_\mathrm{in}$, the lower temperature leads to a decrease in viscosity and hence the outburst enters the decaying phase and the disk inside the ionization front empties on a viscous timescale $t_\mathrm{visc}$, which for our model \texttt{fid} is $t_\mathrm{visc} \lesssim 100$ years in the MRI-active region.

In panel (d) the influence of the protostellar magnetic field (see also \sref{sec:influence_stellar_magnetic_field}) is revealed in the form of a considerably subKeplerian rotational velocity $u_\mathrm{\varphi}$ towards the inner radius. This is due the protostellar magnetic torque acting on the disk. 
As a further consequence, the radial velocity of the disk starts to increase inside the corotation radius $r_\mathrm{cor}$. However, before the radial velocity $u_\mathrm{r}$ is even close to the free-fall velocity, the protostellar magnetic field dominates the disk dynamics \citep[e.g.,][]{bessolaz08} and the accretion flow transitions into accreting funnels. 
This transition zone at the inner rim leads to a drop in surface density $\Sigma$ and as a consequence yields a lower gas temperature $T_\mathrm{gas,0}$. Additionally, because of the decreasing accretion rate $\dot M_\star$, the inner radius is moving outwards (cp.~panels (a) to (d) in \fig{fig:burst_fid_full}), which in combination with the lower temperature leads to an additional cooling wave starting at the inner edge and propagating outwards (color-coded vertical dashed lines and corresponding radially outward-oriented arrows close to the inner rim in panel (b) of \fig{fig:burst_fid_full}). At $t_\mathrm{burst,5}$, the two waves have almost met each other (cp. color-coded vertical lines corresponding to $t_\mathrm{burst,5}$ in panel (b) of \fig{fig:burst_fid_full}), resulting in the turnoff of the TI almost everywhere in the inner region, eventually marking the end of the outburst. Hence, compared to models without the influence of a stellar magnetic field, an additional cooling wave traveling outwards has an influence on the time needed for the inner disk to become thermally stable again. This because the outer cooling front does not travel inwards until it reaches the inner disk rim but is met by the outward-bound cooling wave earlier. Material is then fed to the inner regions from the outer regions and accumulates until a disk annulus becomes thermally unstable again, therefore completing a burst cycle.

\begin{figure}[ht]
    \centering
    \resizebox{\hsize}{!}{\includegraphics{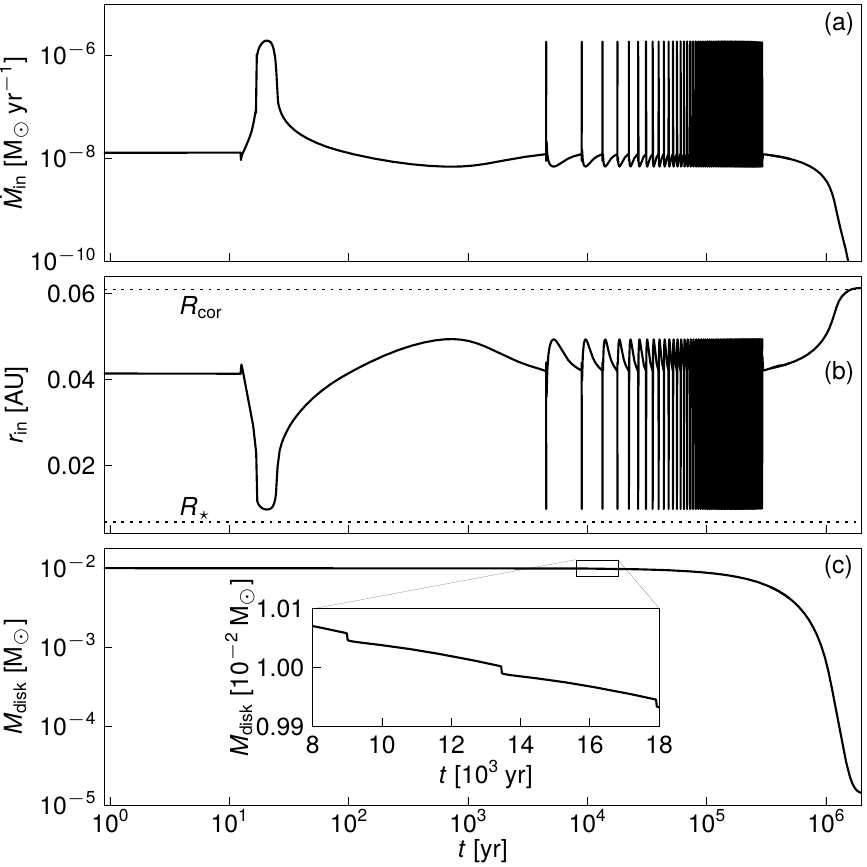}}
    \caption{Long-term evolution for our fiducial model \texttt{fid} (see~\tab{tab:model_parameters}). Panels (a) and (b) show the accretion rate $M_\mathrm{in}$ at the inner boundary and its corresponding movement of the inner radius $r_\mathrm{in}$. The evolving total disk mass $M_\mathrm{disk}$ is presented in panel (c). The inset in panel (c) shows the step-like decrease in disk mass $M_\mathrm{disk}$ caused by outbursts.}
    \label{fig:fiducial_longterm_evolution}
\end{figure}

Such burst cycles occur as long as the inner disk can be fed with material from the outer regions and consequently is capable of heating up the inner regions sufficiently through viscous dissipation. In \fig{fig:fiducial_longterm_evolution} this repeated episodic accretion can be seen in panel (a). This process is repeated $49$ times before the disk mass can no longer provide a sufficiently strong accretion flow of gas to heat the inner disk above the MRI activation temperature $T_\mathrm{active}$ once more. For our fiducial model \texttt{fid}, the disk becomes quiescent at $\sim 2.35 \cdot 10^5$ years and below a disk mass of $\sim 7 \cdot 10^{-3} \, M_\odot$.  Panel (b) of \fig{fig:fiducial_longterm_evolution} shows the variation of the inner radius $r_\mathrm{in}$  as it reacts to a changing accretion rate. After the bursts have ceased and the disk has become quiescent, the accretion rate decreases, as the disk continually loses mass over its lifetime (cp.~panel (c) of \fig{fig:fiducial_longterm_evolution}). The constantly  decreasing accretion rate also has an impact on the location of the inner radius. The magnetic field is able to disrupt the disk farther out, as $\dot M_\star$ becomes weaker. The apparent push of $r_\mathrm{in}$ outside the initial corotation radius $r_\mathrm{cor,init}$ can be explained with a super-Keplerian rotation velocity caused by the positive pressure gradient in the innermost disk in the vicinity of $r_\mathrm{cor,init}$ (cp.~\fig{fig:fiducial_burst_forces}) and therefore a slight change in $r_\mathrm{cor}$ towards larger radii.

\subsection{Influence of the stellar magnetic field}
\label{sec:influence_stellar_magnetic_field}

In \sref{sec:episodic_accretion} we analyze and describe our fiducial model \texttt{fid}. It can be clearly seen that a protostellar magnetic field is influencing the disk close to the inner boundary $r_\mathrm{in}$. In agreement with \citet{Vorobyov19}, we argue that the inner boundary has an effect on the global disk structure and therefore on the long-term evolution of a protoplanetary disk. We emphasize in \sref{sec:1D_approach} that a 1D approach outside of $r \approx 0.5$ AU has certain limitations in describing the disk structure appropriately. However, our focus in this section is to investigate how different magnetic field strengths (see~\tab{tab:model_parameters}) are effecting the bursting behavior of the 
disk.
The MRI and TI ignition point lies well below $0.5$ AU, which is an essentially axisymmetric region of the disk \citep[e.g.,][]{Lesur2020}, and therefore the assumptions of our model are valid approximations. Additionally, we restrict our studies in this work to low-mass disks where no gravitational instabilities are expected to break their axisymmetry.

\begin{figure}[ht]
    \centering
    \resizebox{\hsize}{!}{\includegraphics{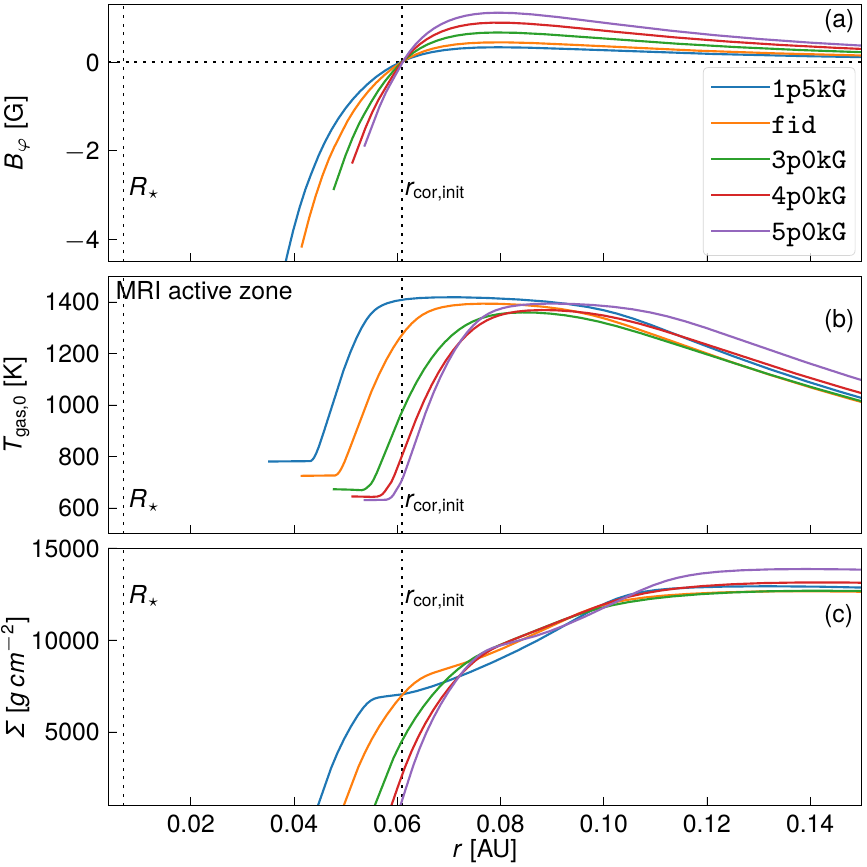}}
    \caption{Comparison of (a) the toroidal protostellar magnetic field component $B_\mathrm{\varphi}$ as stated in \equ{eq:stellar_Bphi}, (b) the midplane gas temperature $T_\mathrm{gas,0}$ , and (c) the surface density $\Sigma$ for the different stationary initial models in \tab{tab:model_parameters} of varying protostellar magnetic field strengths, ranging from $B_\star = 1.5$ kG to $5$ kG. The MRI active zone starts at little over 1400~K for our choice of $T_\mathrm{active} = 1530$~K and $T_\mathrm{width} = 50$~K. The vertical dashed lines denote the position of the stellar radius $R_\star$ and the initial corotation radius $r_\mathrm{cor,init}$.}
    \label{fig:fiducial_cmp_bphi}
\end{figure}

In \fig{fig:fiducial_cmp_bphi} the initial magnetic field strengths for different stationary models (cp.~\tab{tab:model_parameters}) are shown. 
We can determine two effects of the stellar magnetic field. First, the inner boundary of the disk is pushed outward for a higher field strength due to the stronger magnetic pressure (see \equ{eq:truncation_radius}). Additionally, a stronger field also has a stronger propelling effect on the disk gas just outside of the corotation radius $r_\mathrm{cor}$ (cp.~panel (a) of \fig{fig:fiducial_cmp_bphi}). Material in this area starts to get pushed outwards, and the disk starts to pile up material because disk gas is still accreting from the outer disk inwards (see~panel (c) of \fig{fig:fiducial_cmp_bphi}). This leads to increased viscous heating and therefore to a higher midplane gas temperature $T_\mathrm{gas,0}$ (see panel (b) of \fig{fig:fiducial_cmp_bphi}). If the magnetic field is chosen to be stronger than a certain threshold field strength (the exact value depends on the disk and stellar parameters), then the aforementioned pile-up of material is sufficient to trigger the MRI. For the disk parameters chosen in this work, this limiting value is $B_\mathrm{\star,up} \approx 4.5$ kG. Consequently, the higher $T_\mathrm{gas,0}$ is sufficient for deeper disk layers to be become MRI-active, which in turn leads to even higher viscosity and temperature (cp. panel (b) of \fig{fig:fiducial_cmp_bphi}). Eventually, this triggers TI at some ignition radius, which then starts an outburst cycle.

Also notable is the effect of a relatively weak magnetic field on the disk. A lower magnetic field strength yields an inner radius $r_\mathrm{in}$ closer to the protostar. Consequently, the disk is heated up to higher temperatures and the MRI/TI can be triggered. As a result, for all stellar magnetic fields $B_\star$ lower than a lower limit field strength $B_\mathrm{\star,low}$, a disk also undergoes episodic accretion. For the parameters in \tab{tab:model_parameters}, this lower limit is at $B_\mathrm{\star,low} \approx 2$ kG, which we use for our fiducial model \texttt{fid} (cp.~panel (b) of \fig{fig:fiducial_cmp_bphi}).

\begin{figure}[ht]
    \centering
    \resizebox{\hsize}{!}{\includegraphics{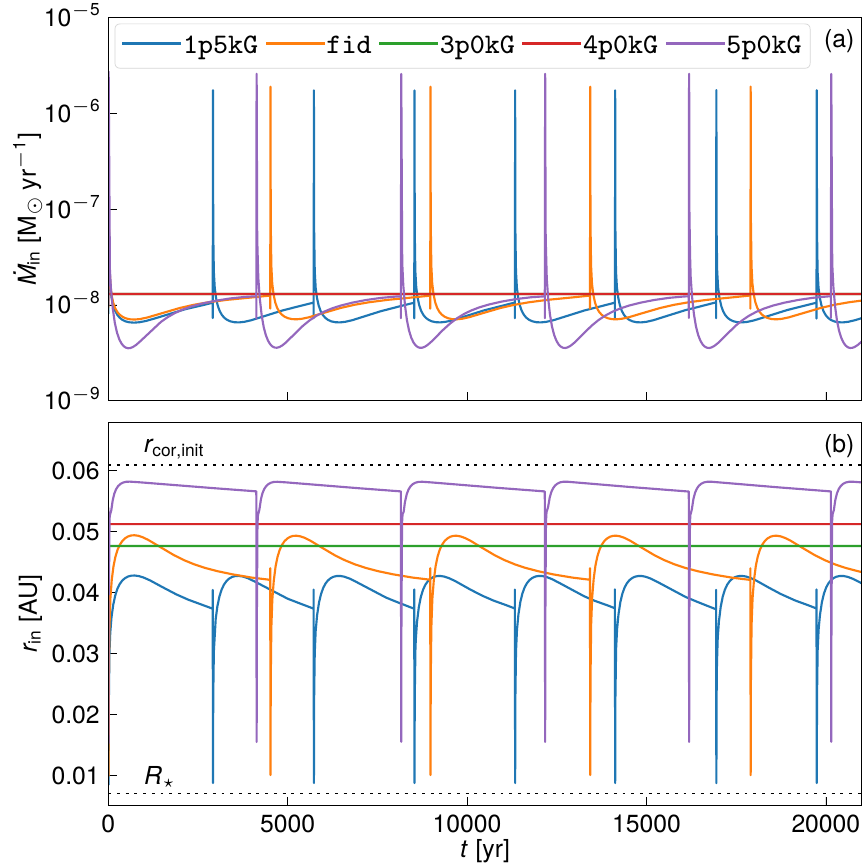}}
    \caption{Comparison of the models in \tab{tab:model_parameters}. Panel (a) shows the accretion rate $\dot M_\star$ and (b) the corresponding inner radius $r_\mathrm{in}$, respectively. The colors correspond to the same models as in \fig{fig:fiducial_cmp_bphi}. The horizontal dashed lines denote the position of the stellar radius $R_\star$ and the initial corotation radius $r_\mathrm{cor,init}$.}
    \label{fig:fiducial_cmp_longterm_detail}
\end{figure}

We also conduct long-term runs for all models in \tab{tab:model_parameters}. In \fig{fig:fiducial_cmp_longterm_detail}, the accretion rate on the star $\dot M_\star$ and the movement of the inner radius $r_\mathrm{in}$ are plotted in panels (a) and (b), respectively. The models \texttt{1p5kG} (blue solid) and \texttt{fid} (orange solid) become thermally unstable due to the proximity of the inner radius $r_\mathrm{in}$ to the star, which can be seen in panel (b) of \fig{fig:fiducial_cmp_longterm_detail}. The burst frequency (cp.~panel (a) of \fig{fig:fiducial_cmp_longterm_detail}) depends on how fast the gas temperature $T_\mathrm{gas,0}$ exceeds the activation temperature $T_\mathrm{active}$. For disks with stellar magnetic field strengths $B_\star < B_\mathrm{\star,low}$ the burst frequency increases for weaker magnetic fields $B_\star$, because the weaker the stellar field $B_\star$, the further inwards the disk stretches. This leads to even higher gas temperatures and therefore to a disk prone to TI. Equally importantly, for $B_\star > B_\mathrm{\star,up}$, the stronger $B_\star$, the more effectively the propelling effect can fling material outwards, and hence the more material piles up outside of $r_\mathrm{cor}$. The gas temperature then also increases with higher $B_\star$ and TI is triggered faster. Models \texttt{1p5kG} and \texttt{fid} satisfy the constraint $B_\star < B_\mathrm{\star,low}$ and \texttt{5p0kG} satisfies $B_\star > B_\mathrm{\star,up}$. For intermediate magnetic fields of $B_\mathrm{\star,low} < B_\star < B_\mathrm{\star,up}$, no outbursting behavior develops (models \texttt{3p0kG} and \texttt{4p0kG}). We note that the actual threshold values for $B_\star$ depend on the choice of stellar and disk parameters and have to be determined anew for every different parameter set  (cp.~\fig{fig:varTactive}). For weaker fields, the inner disk radius $r_\mathrm{in}$ is pushed very close to the stellar radius (cp.~\equ{eq:truncation_radius} and panel~(b)~of~\fig{fig:fiducial_cmp_longterm_detail}); in the case of our model \texttt{1p5kG,} the inner radius $r_\mathrm{in} \approx 1.5\, R_\star$.

\begin{figure}[ht]
    \centering
    \resizebox{\hsize}{!}{\includegraphics{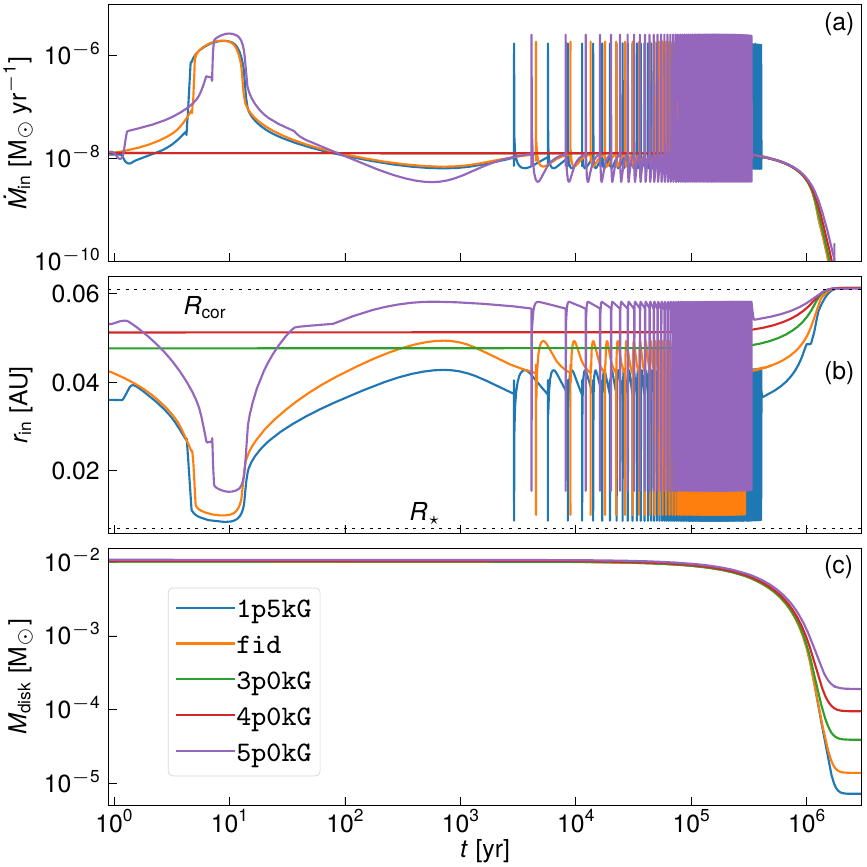}}
    \caption{Same models as in \fig{fig:fiducial_cmp_longterm_detail}, but for the full disk lifetime. The onset of the first burst of every model has been aligned in time to provide a better comparability regarding burst frequency and burst strength. In panel (b) the dashed lines denote the initial corotation radius $R_\mathrm{cor,init}$ and the protostellar radius $R_\star$.}
    \label{fig:fiducial_cmp_longterm}
\end{figure}

In \fig{fig:fiducial_cmp_longterm} the full disk lifetime is plotted. In panels (a) and (b) it can be seen that the transition of the disk into its quiescent state occurs later for models \texttt{1p5kG} and \texttt{5p0kG}, which is due to their higher maximum temperature $T_\mathrm{gas,0}$ (compared to \texttt{fid}) which can also be seen in panel (b) of \fig{fig:fiducial_cmp_bphi}. The physical explanation for the different rest disk masses in (c) of \fig{fig:fiducial_cmp_longterm} is that below a certain disk mass threshold, the protostellar field is effectively preventing further accretion onto the protostar (propeller regime) and hence the disk mass stabilizes at a certain value. The strength of the magnetic field then determines the lower disk mass $M_\mathrm{disk}$, below which further accretion is quenched by $B_\mathrm{\varphi}$.

Towards the end of the disk lifetime, the disk mass and consequently the accretion rate $\dot M_\star$ drop, resulting in $r_\mathrm{in}$ shifting towards larger radii (cp.~\fig{fig:fiducial_cmp_longterm} and \equ{eq:truncation_radius}). With decreasing $\dot M_\star$ the gas pressure  $P_\mathrm{gas}$ eventually exceeds the ram pressure $P_\mathrm{ram}$ and controls the movement of the inner disk radius $r_\mathrm{in}$. For all models, this transition occurs inside the corotation radius $r_\mathrm{cor}$. 
For a sufficiently strong stellar magnetic field, the propelling effect just outside $r_\mathrm{cor}$ piles up material (cp.~\fig{fig:fiducial_cmp_bphi}) and increases $P_\mathrm{gas}$ until it exceeds $P_\mathrm{ram}$,  thus keeping $r_\mathrm{in}$ inside $r_\mathrm{cor}$. 
We want to note here that current 2D simulations of the propelling regime show a more complex behavior \citep[e.g.,][]{Ustyugova06, Romanova09, Romanova18}. A specific part of the piled up matter is accreted onto the star while another is ejected in winds. Inclusion of disk winds in our model, for example, which are shown to play an important role for the inner boundary \citep[e.g.,][]{Koenigl11},  would reduce pile up of matter (see panel (a) in \fig{fig:fiducial_cmp_bphi}) and consequently increase $B_\mathrm{_\star,up}$. Thus, the results presented here have to be considered as a limiting case.

\begin{figure}[ht]
    \centering
    \resizebox{\hsize}{!}{\includegraphics{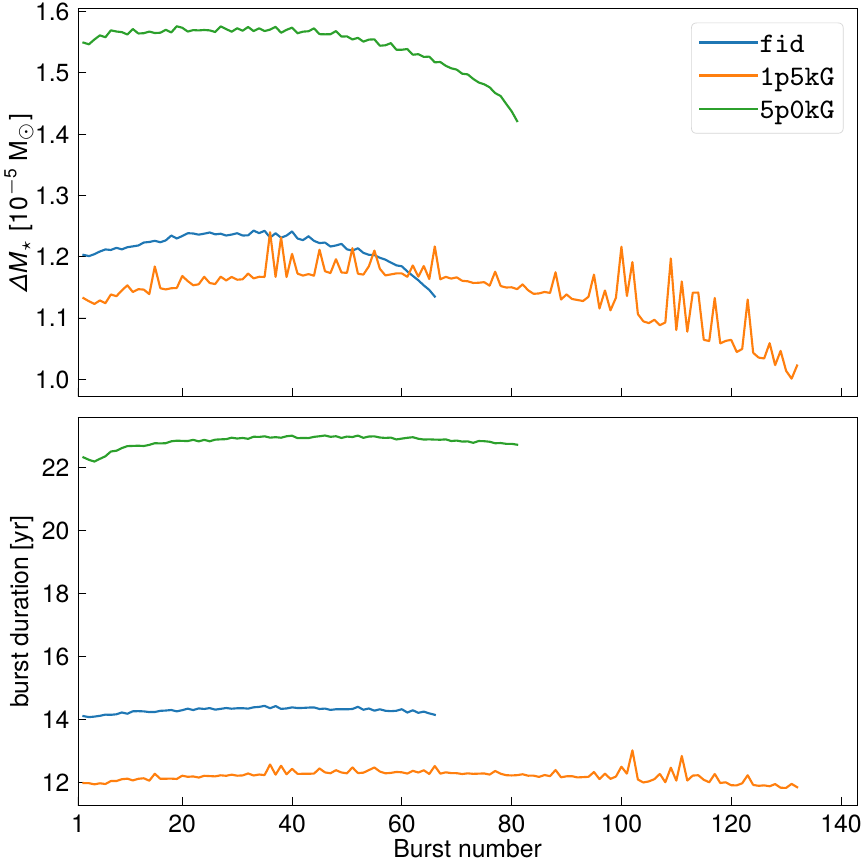}}
    \caption{Time evolution of the outbursting behavior  of the disk, shown for all consecutive bursts during disk evolution. Panel (a) shows the mass accretion onto the protostar during an outburst $\Delta M_\star$, whereas panel (b) depicts the burst duration for a certain burst.}
    \label{fig:analysis_plot}
\end{figure}

In addition to the long-term evolution of the disk for different values of the stellar magnetic field, we compare the duration of bursts and the accreted disk mass onto the star $\Delta M_\star$ for each individual outburst (cp.~\fig{fig:analysis_plot}). For a given stellar magnetic field strength, the burst duration as well as $\Delta M_\star$ remain approximately constant until the disk has depleted to a degree where the stellar disk can no longer feed the inner region with sufficient mass  to become MRI- or TI-unstable. The small variability of the duration and the accreted mass of consecutive bursts is due to the fact that the ignition radius for the MRI/TI remains effectively unaltered for the disk during its bursting phase, which then yields a very similar maximum radius for the outward-bound ionization front and consequently a similar burst duration and $\Delta M_\star$. Relaxing the assumption of axisymmetry probably adds more variability to the results \citep[e.g.,][]{vorobyov20} and shall be reviewed in further studies. However, for an increasing stellar magnetic field strength, the burst duration and $\Delta M_\star$ increase. Because of the stronger magnetic pressure with an increasing stellar magnetic field, the inner disk radius $r_\mathrm{in}$ as well as the MRI ignition point are pushed outwards (cp.~panel~(b)~of~\fig{fig:fiducial_cmp_bphi}). Additionally, more mass is piled up directly outside the MRI ignition point (cp. panel (c) of \fig{fig:fiducial_cmp_bphi}), which also increases the disk temperature in this region. As a result, the outward-bound ionization front reaches further out, which increases the burst duration and the piled up mass outside the MRI ignition point increases $\Delta M_\star$.

\begin{figure}[ht]
    \centering
    \resizebox{\hsize}{!}{\includegraphics{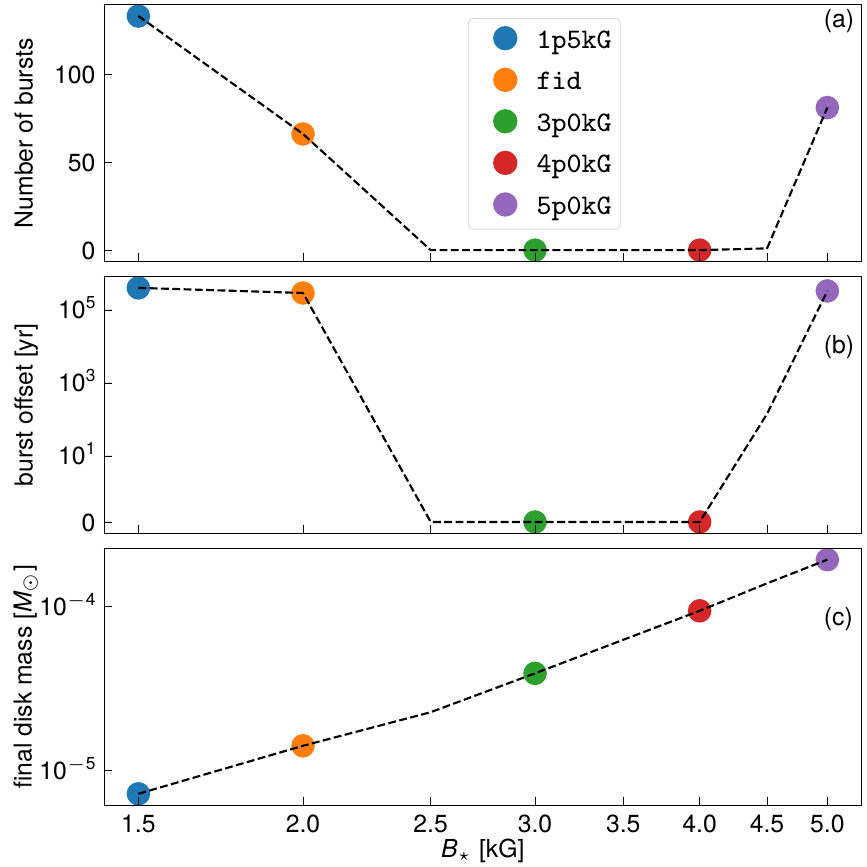}}
    \caption{ Comparison of (a) the number of bursts, (b) the burst offset time of the disk, and (c) the final disk mass for all models in \tab{tab:model_parameters} and additionally for models with $B_\star = 2.5$ kG, $3.5$ kG and $4.5$ kG. The color coding is the same as for \fig{fig:fiducial_cmp_bphi}, whereas the black dashed line linearly connects the data points and has the sole purpose of emphasizing the burst-free region in panels (a) and (b) as well as indicating the power-law for the final mass in (c).}
    \label{fig:fiducial_cmp_stats}
\end{figure}

\figl{fig:fiducial_cmp_stats} specifically depicts the thermally stable zone for $B_\mathrm{\star}$ between roughly $2.5$ kG and $4.5$ kG. The final disk mass towards the end of a disk's lifetime seems to behave like a power-law (cp.~panel (c) of \fig{fig:fiducial_cmp_stats}). A possible explanation for this is that an almost dissolved disk with little mass cannot counteract the propelling forces of the stellar magnetic field and hence cannot accrete onto the star. The stellar field is modeled as a dipole and the resulting centrifugal force would also yield a power law; nevertheless a more thorough analysis is planned in further studies to investigate this behavior in more detail.
Panel (b) of \fig{fig:fiducial_cmp_stats} depicts the point in time where the disk ceases to burst (burst offset), whereas panel (a) shows the number of bursts for a certain model. \fig{fig:fiducial_cmp_stats} indicates that stellar magnetic fields, which are weaker than the lower limit or stronger than the upper limit of $B_\star$, will lead to more bursts, depending on the size of the deviations from those limits.

\subsection{Variation of activation temperature and mass transport rate}
\label{sec:influence_tcrit}

\begin{figure}[ht]
    \centering
    \resizebox{\hsize}{!}{\includegraphics{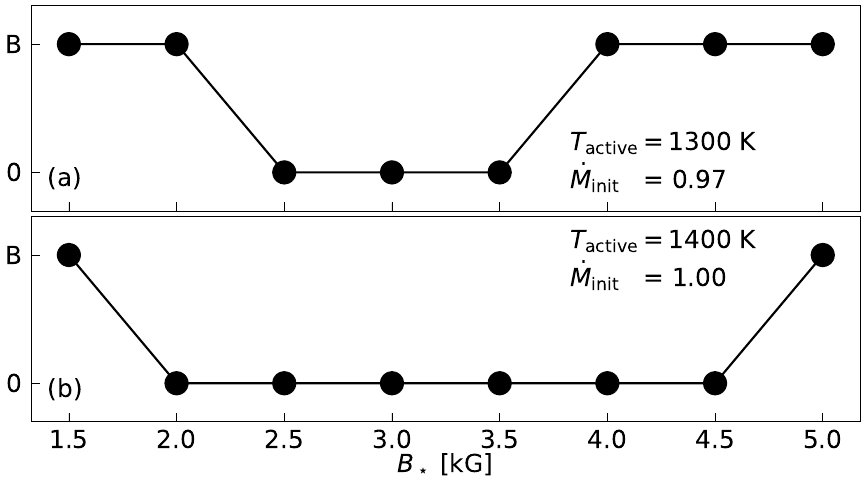}}
    \caption{Bursting behavior depending on the stellar magnetic field $B_\star$ for different $T_\mathrm{active}$. A distinction is made between models that produce outbursts (B) and models in which outbursts are suppressed ($0$).  $T_\mathrm{active}$ is set to 1300 K and 1400 K in panels (a)  and (b), respectively. The accretion rate $\dot{M}_\mathrm{init}$ is given in units of $10^{-8} \mathrm{M_\odot / yr^{-1}}$.}
    \label{fig:varTactive}
\end{figure}

To test the robustness of the results of \sref{sec:influence_stellar_magnetic_field}, we varied the MRI activation temperature $T_\mathrm{active}$ together with the initial mass transport rate $\dot M_\mathrm{init}$ . For all previous simulations, $T_\mathrm{active}$ is fixed to 1530~K (see~\tab{tab:model_parameters}). A different choice of $T_\mathrm{active}$ will affect the radius at which the deep layer of the  disk becomes MRI-active and TI is triggered. Choosing a lower activation temperature will result in more models becoming thermally unstable. The value range between $B_\mathrm{\star,low}$ and $B_\mathrm{\star,up}$ becomes smaller. A higher activation temperature $T_\mathrm{active}$, on the other hand, will increase this region as there will be fewer models in which  $T_\mathrm{gas}$  can exceed $T_\mathrm{active}$. We can reproduce our main results for different values of $T_\mathrm{active}$ by adapting the mass transport rate $\dot M$ throughout the disk. Choosing $T_\mathrm{active} = 1300$~K and $\dot{M}_\mathrm{init}$ of $0.97 \cdot 10^{-8} \mathrm{M_\odot / yr^{-1}}$, the range of the stellar magnetic field $B_\star$ in which outbursts are suppressed extends from 2.5 to 3.5~kG (see \fig{fig:varTactive}). For choosing $T_\mathrm{active} = 1400$~K and $\dot{M}_\mathrm{init}$ of $1.0 \cdot 10^{-8} \mathrm{M_\odot / yr^{-1}}$, outbursts are suppressed between 2.0 and 4.5~kG. The adapted values of $\dot{M}_\mathrm{init}$, which correspond to the quiescent accretion rate between two outbursts, are all in agreement with \citet{armitage01}. We would like to emphasize that the exact interpretation of these results requires a more detailed parameter study and is left for future work.

\section{Conclusion}
\label{sec:conclusion}

In this study we conduct fully implicit 1+1D hydrodynamic simulations of protoplanetary disks with a focus on the inner regions. The TAPIR code \citep[see][]{ragossnig20} is capable of solving the full set of hydrodynamic equations, which enables us to include additional physics, such as for example the effects of local pressure gradients, protostellar magnetic fields, or features in the angular velocity. 

Previous studies \citep[e.g.,][]{bell94, armitage01, zhu09} wherein 1D long-term simulations were also performed used the viscous diffusion equation for the temporal evolution of the disk. We demonstrate in \sref{sec:inner_boundary} that this approach shows significant deviations ---at least locally close to the star--- from a full solution of the hydrodynamic equations. Furthermore, we also adopt a description of the magnetic truncation radius based on the equality of magnetic pressure and ram pressure \citep[see e.g.,][]{hartmann16}. Complementary to that approach, we incorporate the gas pressure influence in our models, covering $P_\mathrm{gas} > P_\mathrm{ram}$.

In the first paper of this series, we study the effect of protostellar magnetic torques on the long-term behavior of protoplanetary disks. These torques slow down or accelerate the disk depending on the radial location compared to the corotation radius $r_\mathrm{cor}$, and thus necessitate a self-consistent treatment of the momentum equation in the angular direction. Diffusion approaches are not well suited to modeling an inner disk threaded by a stellar magnetic field.

Layered-viscosity models as proposed by \citet{armitage01} can, for certain parameters, become thermally unstable and repeatedly experience FU Ori-like episodic accretion events \citep[e.g.,][]{bell94}. Interestingly, the burst strength and burst frequency are strongly dependent on the stellar magnetic field strength $B_\star$.

On the one hand, a stronger $B_\star$ corresponds to a larger inner disk radius $r_\mathrm{in}$, which leads to less radiative heating of the surface of the  inner
disk and consequently to a cooler disk. On the other hand, a stronger toroidal magnetic field component $B_\mathrm{\varphi}$ also acts as a propeller outside of the corotation radius $r_\mathrm{cor}$, effectively flinging material outwards. This pushed out disk gas combines with the inward-directed accretion flow, resulting in a larger  radially localized surface density $\Sigma$. This leads to a higher gas temperature $T_\mathrm{gas}$, which counteracts the cooling effect of a farther out $r_\mathrm{in}$. We find that there exists a limit value for $B_\star$, above which a disk gets sufficiently heated to become thermally unstable.

A weaker $B_\star$ leads to a small inner radius $r_\mathrm{in}$, and therefore the inner disk gets hotter due to the proximity to the protostar. We find that if the protostellar magnetic field $B_\star$ is weaker than a certain upper limit, then the inner disk becomes hot enough to also become thermally unstable.

Summarizing these results, a region of  protostellar magnetic field of intermediate strength exists where (for a certain set of disk and star parameters) no bursts occur. Below and above this region the models can become thermally unstable. These results are later reproduced with different value sets of the MRI activation temperature $T_\mathrm{active}$ and mass transport rate $\dot M$. Additionally, the burst strength and burst frequency are also altered by a varying $B_\star$, which changes the duration of the thermally unstable phase of such disks. We find that the burst duration and mass accretion rate caused by a single outburst only vary slightly for consecutive bursts, which is due to the almost constant radial position of the MRI/TI ignition point during the bursting phase of a disk. Moreover, for stronger stellar magnetic fields, the burst duration and accretion mass during an outburst increase because the MRI ignition point gets pushed radially outwards.
We conducted long-term simulations to study the  influence of the protostellar field and found that disks tend to have more frequent but weaker bursts if $B_\star$ is either far below or far above the aforementioned intermediate zone. Our simulations show that for weaker stellar magnetic fields, the inner disk radius $r_\mathrm{in}$ is pushed very close to the stellar radius $R_\star$. For even weaker fields or stronger bursts the magnetic field could be squashed to the stellar surface and could therefore suppress the accretion funnels as well as the resulting accretion shock \citep[e.g.,][]{hartmann16} and thus provide an explanation for the lack of UV excess in FU Ori-like star--disk systems \citep[e.g.,][]{bell94}.

In this work, we show that a detailed model of the inner disk regions is essential for the simulations of protoplanetary disks, especially during episodic outbursts. We agree in this regard with  \citet{Vorobyov19}. for example, and are confident that the TAPIR code \citep[][]{ragossnig20} is well suited to representing the inner disk region in global 2D long-term simulations.

\subsection{Limitations of this work}

While the TAPIR-code is capable of treating the inner disk self-consistently, there are also limitations to this approach. Our 1D model is mainly suited to simulation of the inner disk. This is because in the inner regions, viscous shear becomes stronger and therefore angular perturbations cannot be sustained, even for short times \citep[for a conclusive discussion see e.g.,][]{Lesur2020}. Therefore the assumption of axisymmetry is a valid approximation in the inner regions.

Furthermore, because of the assumption of vertical thermal equilibrium, radiative transport in the z-direction is only included approximately. 
Photoevaporation of the inner disk definitely has an effect on the inner radius of the disk, as well as on the star--disk-interaction as gas is removed, especially from the inner part of the disk. However, these effects will not effect the qualitative statements of this work and will be investigated in upcoming studies.
Furthermore, we use a simple dust model with a constant dust-to-gas ratio. The dust is included in the opacity calculation, but the back-reaction onto the gas or dust dynamics is not considered yet. This approximation has only a minor effect on the main findings from this study. However, a more realistic time-dependent dust model would increase the accuracy of our code.

We also neglect some details of the star--disk interaction. In our model, $B_\star$ is prescribed as a dipole field, which is an approximation, especially in the case of very high accretion rates. Nevertheless, for an qualitative study of the effects of magnetic torque on the disk evolution, the details of the field topology are of minor order. Additionally, the transfer of angular momentum from the star to the disk and vice versa has not been modeled in this work, as we focused on the effect of an existing magnetic field on disk evolution. We agree that this effect should be investigated in further studies, because a time-dependent stellar angular velocity also results in the corotation radius $r_\mathrm{cor}$ moving radially inwards and outwards, which has a potential effect on the outbursting behavior of disks.

\subsection{Further work}

In upcoming studies of this series, we will study the effects of magnetic fields and stellar radiation on the pressure scale height, which leads to a shadowed disk region, where less stellar radiation can heat the disk surface.
Another study will involve a model for the star--disk interaction and the inclusion of higher-order magnetic field moments like quadrupole and octopole contributions to the stellar magnetic field topology. We will use the altered magnetic field structure to study the effect on the outbursting behavior of disks.

Equally interesting is the inclusion of large-scale disk fields. Those fields have been investigated in 1D models \citep[e.g.,][]{Guilet2012,Guilet2014}. The incorporation of such magnetic field models in the TAPIR code enables us to self-consistently solve the inner disk and the disk field topology simultaneously. This further allows estimates of how much mass and angular momentum can be transported away by magneto-centrifugally driven outflows during for example an FU Ori-like outburst. Also important is the inclusion of photo-evaporative winds, especially during an outburst, because  this is when the accretion luminosity is highly elevated. The effects of outflows due to photo-evaporation coupling with a large-scale disk magnetic field are equally interesting, as then not only mass is transported away but also angular momentum, which further modifies the disk dynamics. 

\begin{acknowledgements}
  E.I.V. acknowledge support of Ministry of Science and Higher Education of the Russian Federation under the grant 075-15-2020-780 (N13.1902.21.0039; Section 6).
 \end{acknowledgements}

\appendix
\section{Implicit time-step}
\label{sec:imp_timestep}
\begin{figure}[h]
    \centering
         \resizebox{\hsize}{!}{\includegraphics{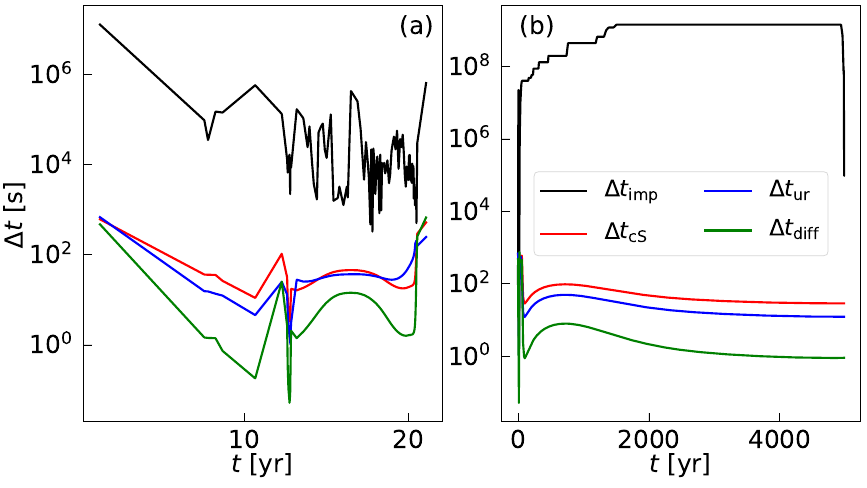}}
    \caption{Comparison of the implicit time-step $\Delta t_\mathrm{imp}$ to time-step limitations imposed by the CFL condition with respect to the speed of sound $\Delta t_\mathrm{cS}$, the radial velocity $\Delta t_\mathrm{ur}$, and the diffusion $\Delta t_\mathrm{diff}$ for our fiducial model; (a) during an outburst and (b) during the quiescent phase between two outbursts. The decrease in $\Delta t_\mathrm{imp}$ at the right side denotes the beginning of the next outburst.}
    \label{fig:implicit_timestep}
\end{figure}

Implicit numerical models are not bound to the CFL condition and can benefit from larger time-steps and consequently shorter simulation times compared to explicit models. To quantify this difference, we compare the maximum possible time-ste--- $\Delta t \leq \Delta r / u_\mathrm{i}$ ---allowed by the CFL condition; with the radial grid size $\Delta r$ and a velocity $u_\mathrm{i}$ with which information propagates. In an accretion disk, there are several velocities that transport information. For this order-of-magnitude approximation, we want to compare the effects of the speed of sound $c_\mathrm{S}$ with those of the radial velocity $u_\mathrm{r}$. Additionally, the time-step limitation of diffusion in the disk--- $\Delta t \leq \Delta r^2 / D$ ---is taken into account. Here, $D$ is the diffusion coefficient and is connected to the viscosity $\nu$ via the Schmidt number $Sc = \nu / D$. Following \cite{armitage11}, $Sc$ is of the order of unity and $Sc = 1$ is used. In \fig{fig:implicit_timestep}, the time-step used in the implicit TAPIR code is compared to the theoretical time-step limitations imposed by the CFL condition for an explicit method. During an outburst, the implicit time-step exceeds the explicit time-step by up to two orders of magnitude and during the quiescent phase, by up to six orders of magnitude. Our model also incorporates an adaptive time-step control, which increases or decreases $\Delta t$ until the Newton-Raphson iteration (cp.~\sref{sec:implicit_integration}) converges towards a solution. This results in shorter time-steps during bursts and longer time-steps in quiet phases, because in the latter case the disk quantities change slowly over time and a solution can still be found for longer time-steps. We can conclude that an explicit model with the same radial resolution would increase the computational time from days to months or even years, which again justifies this implicit approach.

\bibliographystyle{resources/bibtex/aa}
\bibliography{ppd_inner_boundary_final}

\end{document}